\documentclass[%
 aip,
 amsmath,amssymb,
 reprint,%
]{revtex4-1}

\usepackage{graphicx}
\usepackage{dcolumn}
\usepackage{bm}

\usepackage[utf8]{inputenc}
\usepackage[T1]{fontenc}
\usepackage{mathptmx}
\usepackage{etoolbox}
\usepackage{soul}
\usepackage{xcolor}
\usepackage{siunitx}
\usepackage{multirow}
\makeatletter
\def\@email#1#2{%
 \endgroup
 \patchcmd{\titleblock@produce}
  {\frontmatter@RRAPformat}
  {\frontmatter@RRAPformat{\produce@RRAP{*#1\href{mailto:#2}{#2}}}\frontmatter@RRAPformat}
  {}{}
}%
\makeatother
\begin{document}

\preprint{AIP/123-QED}

\title[Revealing epitaxial relationships at Ga$_2$O$_3$ interfaces with \textit{p}-type oxides.]{Revealing epitaxial relationships at Ga$_2$O$_3$ interfaces with \textit{p}-type oxides.}

\author{Anna Sacchi}
 \email{anna.sacchi@nlr.gov}
 \affiliation{Materials Science Center, National Laboratory of the Rockies, Golden, CO, USA}
 
 \author{Krishna Acharya}
 \affiliation{Metallurgical and Materials Engineering Department, Colorado School of Mines, Golden, CO, USA}

 \author{Michelle A. Smeaton}
  \affiliation{Materials Science Center, National Laboratory of the Rockies, Golden, CO, USA}

\author{Renae N. Gannon }
  \affiliation{Materials Science Center, National Laboratory of the Rockies, Golden, CO, USA}
  
\author{Vladan Stevanovic}
\affiliation{Metallurgical and Materials Engineering Department, Colorado School of Mines, Golden, CO, USA}

 \author{Steven R. Spurgeon}
\affiliation{Materials Science Center, National Laboratory of the Rockies, Golden, CO, USA}
\affiliation{Metallurgical and Materials Engineering Department, Colorado School of Mines, Golden, CO, USA}
\affiliation{Renewable and Sustainable Energy Institute, University of Colorado Boulder, Boulder, CO, USA}

\author{M. Brooks Tellekamp}
\affiliation{Materials Science Center, National Laboratory of the Rockies, Golden, CO, USA}

\author{Andriy Zakutayev}
 \email{andriy.zakutayev@nlr.gov}
\affiliation{Materials Science Center, National Laboratory of the Rockies, Golden, CO, USA}

\date{\today}

\begin{abstract}
\textit{p}-type oxide contact layers such as Cr$_2$O$_3$ and NiO are attracting increasing interest in \textit{pn}-heterojunctions with \textit{n}-type monoclinic $\beta$-Ga$_2$O$_3$ for high-power electronic devices and other extreme environment applications. However, scientific understanding of their epitaxial relationships remains incomplete. In this work we investigate the epitaxial relation of Cr$_2$O$_3$ and NiO layers to (001) and ($\bar{2}$01) out-of-plane oriented Ga$_2$O$_3$ substrates. Surprisingly, we find that, for the most commercially relevant (001)-orientation of the Ga$_2$O$_3$ substrate, the epitaxial relationships are Cr$_2$O$_3$ (0001) and NiO (111) $\parallel$ Ga$_2$O$_3$ (101), both at the non-intuitive $\chi$ = 22.5 $^\circ$ angle with respect to the substrate normal. We explain this unusual discovery by the interfacial atomistic bonding dominated by oxygen sublattice equivalence of these Cr$_2$O$_3$ and NiO polar surface orientations to the tilted Ga$_2$O$_3$ (101), rather than Ga$_2$O$_3$ (001) substrate surface planes. Furthermore, we assign the in-plane orientation for Cr$_2$O$_3$ on ($\bar{2}$01)-oriented Ga$_2$O$_3$ as: Cr$_2$O$_3$ $[12\bar{3}0]$ $\parallel$ Ga$_2$O$_3$ $[010]$ with two in-plane rotational domains. Interface modeling confirms the in-plane orientation for Cr$_2$O$_3$/$(\bar{2}01)$ Ga$_2$O$_3$ and  shows that strained O-terminated Ga$_2$O$_3$ $(\bar{2}01)$ surfaces have the lowest interfacial energy with Cr$_2$O$_3$ (0001). Beyond establishing the specific epitaxial relationships for Cr$_2$O$_3$ and NiO on Ga$_2$O$_3$, this work provides a systematic methodology for the unambiguous structural characterization of heterointerfaces involving materials with markedly different crystal symmetries.\end{abstract}

\maketitle
In heteroepitaxial growth, crystal substrate  quality, orientation, and symmetry strongly influence epitaxial layer quality, and affects the resulting electrical and thermal transport properties through the interface . This is particularly important for the symmetry-mismatched substrates, where anisotropy can produce complex epitaxial relationships. Examples include GaN/Al$_2$O$_3$, where large lattice constant mismatch generates rotated in-plane domains, leading to high density of threading dislocations;\cite{https://doi.org/10.1002/pssb.200303368} cubic SiGe on hexagonal Al$_2$O$_3$, which locally rotates and distorts to accommodate the trigonal surface, leading to thermoelectric property change;\cite{park_rhombohedral_2008} and $\kappa$-Ga$_2$O$_3$ on GaN or STO, which forms persistent three-fold rotational domains, affecting in-plane transport properties.\cite{nishinaka_microstructures_2018, https://doi.org/10.1002/adfm.202207821} A similar behavior can therefore be expected for hexagonal corundum $\alpha$-Cr$_2$O$_3$ or cubic rocksalt NiO grown on monoclinic $\beta$ phase of gallium oxide ($\beta$-Ga$_2$O$_3$), the system here investigated.\\
Interfaces between \textit{p}-type materials like $\alpha$-Cr$_2$O$_3$ and NiO and $\beta$-Ga$_2$O$_3$ (from now on Ga$_2$O$_3$) are of significant interest for power electronics, owing to the potential of Ga$_2$O$_3$ as an ultrawide bandgap semiconductor, \textit{i.e.}, high breakdown field, tunable \textit{n}-type conductivity, chemically robust.\cite{higashiwaki_-ga2o3_2022} However, the lack of \textit{p}-type conductivity in Ga$_2$O$_3$ requires the development of \textit{p}-type/Ga$_2$O$_3$ heterojunctions, in order to enable bipolar device architectures. Among the different \textit{p}-type materials investigated, \textit{p}-type metal oxides are particularly attractive due to their higher chemical compatibility with Ga$_{2}$O$_{3}$.\cite{kokubun_all-oxide_2016} NiO is to date the most extensively studied, with vertical NiO/Ga$_{2}$O$_{3}$ devices reporting breakdown voltages ranging up to 13.5 kV.\cite{li_breakdown_2024} However, the thermal instability of the interface with the reported formation of an intermixed spinel phase, NiGa$_{2}$O$_{4}$, \cite{egbo_niga2o4_2024,egbo_epitaxial_2025} raises concerns regarding the long-term stability and intrinsic quality of the NiO/Ga$_{2}$O$_{3}$ interface. These concerns are particularly relevant, not only for high-temperature operation, but also in light of local heating due to the intrinsically low thermal conductivity of Ga$_2$O$_3$. Together, these factors motivate a more detailed investigation of the interface structure\cite{smeaton2026revealingatomicstructurenioga2o3} as well as investigation of alternative \textit{p}-types oxides.\\
Corundum $\alpha$-Cr$_{2}$O$_{3}$ (from now on Cr$_2$O$_3$) has gained increasing attention as new candidate for \textit{p}-type contact on Ga$_{2}$O$_{3}$, owing to its potentially higher thermal and environmental stability.\cite{callahan_reliable_2024,liu_electrical_2025} High-performance Cr$_{2}$O$_{3}$/Ga$_{2}$O$_{3}$ diodes have been reported, with breakdown fields reaching up to 12.9 MV/cm; however the breakdown field is highly dependent on the substrate orientation, ranging from 2.7 to 12.9 MV/cm.\cite{liu_cr2o3beta-ga2o3_2025} Despite compelling electrical device reports, the epitaxial relationship and structural quality of the Cr$_{2}$O$_{3}$/Ga$_{2}$O$_{3}$ interface remains poorly understood. Indeed, the interface between the trigonal unit cell of  Cr$_{2}$O$_{3}$, (group symmetry \textit{R$\bar{3}$c}, lattice parameters: \textit{a} = \textit{b} = 4.957$\si{\angstrom}$, \textit{c} = 13.592$\si{\angstrom}$, $\textit{$\alpha$} = \textit{$\beta$} = 90^\circ, \textit{$\gamma$} = 120^\circ$) \cite{sawada_residual_1994} and the monoclinic unit cell of Ga$_{2}$O$_{3}$ (group symmetry \textit{C2/m}, lattice parameters: \textit{a} = 12.226 $\si{\angstrom}$, \textit{b} = 3.041 $\si{\angstrom}$, \textit{c} =  5.809$\si{\angstrom}$, $\textit{$\alpha$} =  \textit{$\gamma$}= 90 ^\circ,\textit{$\beta$} = 103 ^\circ$),\cite{geller_crystal_1960} consists of a high/low structural symmetry interface, making the epitaxial relationship nontrivial. A similar challenge has been encountered for cubic-NiO/monoclinic-Ga$_{2}$O$_{3}$ interfaces.\cite{smeaton2026revealingatomicstructurenioga2o3} The anisotropy of Ga$_{2}$O$_{3}$  monoclinic unit cell complicates the understanding of the interface, as different crystallographic planes/substrate orientations can exhibit distinct structural and physical properties. To date, only three studies have reported structural characterization of Cr$_{2}$O$_{3}$/Ga$_{2}$O$_{3}$.\cite{ghosh_epitaxial_2019, ghosh_evaluation_2021,callahan_reliable_2024}  For both (\={2}01) and (001) oriented Ga$_{2}$O$_{3}$, the reported epitaxial relationship of Cr$_{2}$O$_{3}$ is: Cr$_{2}$O$_{3}$ (0001) || Ga$_{2}$O$_{3}$ (\={2}01), similar to that observed for the similar monoclinic-Ga$_{2}$O$_{3}$/trigonal $\alpha$-Al$_{2}$O$_{3}$.\cite{nakagomi_crystal_2012} However, the corresponding in-plane epitaxial relationship is not obvious and has not been addressed properly.\\

\begin{figure}
\includegraphics[width=\columnwidth]{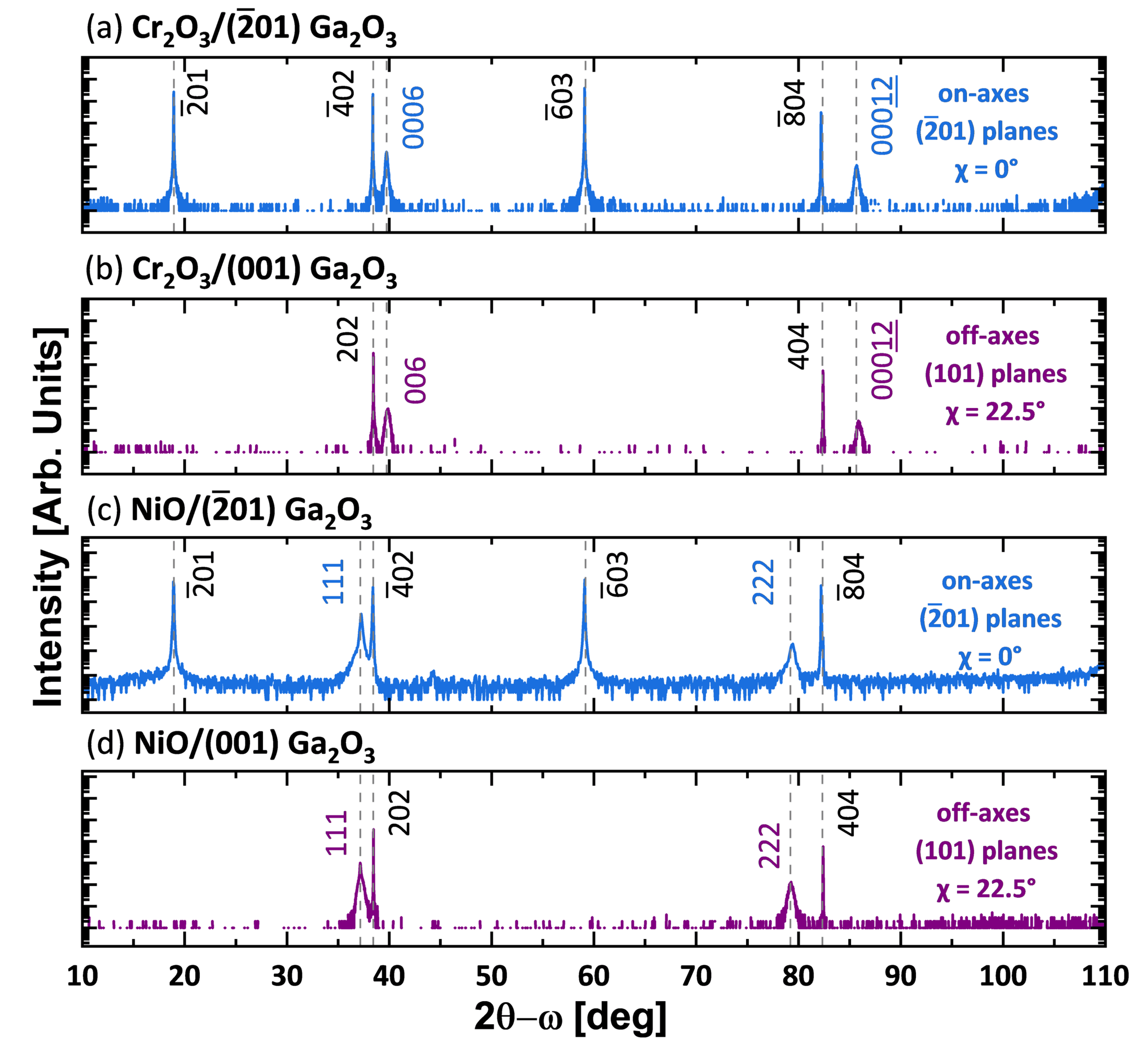}
   \caption{\label{fig:Fig1} In (a,b) high resolution 2$\theta$-$\omega$ scans for Cr$_2$O$_3$ epilayers grown on ($\bar{2}$01) and (001) Ga$_2$O$_3$, acquired at $\chi$ = 0° to probe the out-of-plane scattering vectors and at $\chi$ = 22.5° to probe in-plane (101) Ga$_2$O$_3$, respectively. In (c,d) same analysis is performed for NiO epilayers grown on ($\bar{2}$01) and (001) Ga$_2$O$_3$. Each peak is labeled and vertical dashed line are traced at nominal bulk 2$\theta$ positions, for both substrate and epilayer.}
\end{figure}
In this work, we first identify the epitaxial orientation of Cr$_{2}$O$_{3}$ grown on (001) Ga$_{2}$O$_{3}$, finding off-normal characterization is required and that the  (0001) Cr$_{2}$O$_{3}$ planes align to Ga$_{2}$O$_{3}$ (101). We discuss the origin of the detected epitaxial orientation in terms of epilayer/substrate oxygen anion sublattice equivalence.\cite{oshima_mapping_2026} We determine the in-plane crystallographic orientation of Cr$_2$O$_3$ on ($\bar{2}$01) Ga$_2$O$_3$, and report the presence of in-plane rotational domains. We also report interface-modeling for Cr$_{2}$O$_{3}$/Ga$_{2}$O$_{3}$ system, using a previously developed structure-matching algorithm which confirms our experimental findings.\cite{Therrien_JCP_2020, smeaton2026revealingatomicstructurenioga2o3} Furthermore, we extend the same structural characterization methodology to NiO/(001) Ga$_{2}$O$_{3}$ heterostructures: we find the same off-normal analysis is required to characterize the NiO structure and epitaxial relationship with Ga$_{2}$O$_{3}$. This work enables a direct comparison of the two \textit{p}-type oxide/Ga$_2$O$_3$ systems, while providing a robust methodology for the structural characterization of high/low symmetry heterointerfaces.\\

The Cr$_2$O$_3$ and NiO epilayers investigated in this work are deposited on commercially available $\beta$-Ga$_2$O$_3$ substrates from Novel Crystal Technology, with nominal (001) and (\={2}01) out-of-plane orientation. Prior to deposition, the substrates are cleaned using organic solvents (acetone, methanol, isopropanol) followed by a piranha solution (H$_2$O$_2$:H$_2$SO$_4$ = 1:4), to remove the protective photoresist. The substrates are mounted onto a 2" Si wafer through molten indium, to ensure good thermal contact, and are annealed in the deposition chamber in O$_2$ atmosphere for 10 min at temperature 50 $^\circ$ hotter than the target growth temperature. The epilayers are deposited by pulsed laser deposition (PLD) using a KrF excimer UV laser ($\lambda$ = 248 nm), operated at a pulse energy of 350 mJ and a repetition rate of 10 Hz. Stoichiometric NiO and Cr$_2$O$_3$ targets are ablated in an O$_2$ atmosphere. For NiO, the deposition is performed at an O$_2$ partial pressure of 50 mTorr and a substrate temperature of 300 °C, whereas Cr$_2$O$_3$ is deposited at 18 mTorr and 600 °C.\\
The crystal structure and epitaxial quality of the films are characterized by X-ray diffraction (XRD) using two instruments. A Bruker D8 powder diffractometer, equipped with HI-STAR 2D detector and a Cu K$\alpha$ source, used to acquire symmetric 2$\theta$–$\omega$ scans over a broad range of in-plane $\chi$ angles. This enables simultaneous identification of diffraction features associated with out-of-plane and in-plane orientations. The epilayers are further investigated with a Rigaku SmartLab high resolution (HR) diffractometer, equipped with a Cu K$\alpha$ source, a Ge(220) monochromator, and an Eulerian cradle, which enables access to specific in-plane reflections ($\chi$ $\neq$0°). The experimental epitaxial relationships are then reported schematically using VESTA software.\cite{momma_vesta3_2011} In the Supplementary Information (SI) - Section I, we report for clarity the notation used to describe structural/crystallographic objects throughout the manuscript.\\
First-principles density functional theory (DFT) calculations were performed using the Vienna Ab initio Simulation Package (VASP)\cite{kresse_PRB_1999} to determine the lattice parameters of the material under investigation, \textit{i.e.}, Ga$_2$O$_3$ and Cr$_2$O$_3$, in their bulk shape. Interface modeling is performed by applying the structure-mapping algorithm developed by Therrien \textit{et al.},\cite{Therrien_JCP_2020,Therrien_PRAppl_2021} utilizing the open-source Python package \textit{p2ptrans}.\cite{p2ptrans_2022}

\begin{figure}[h]
\includegraphics[width=\columnwidth]{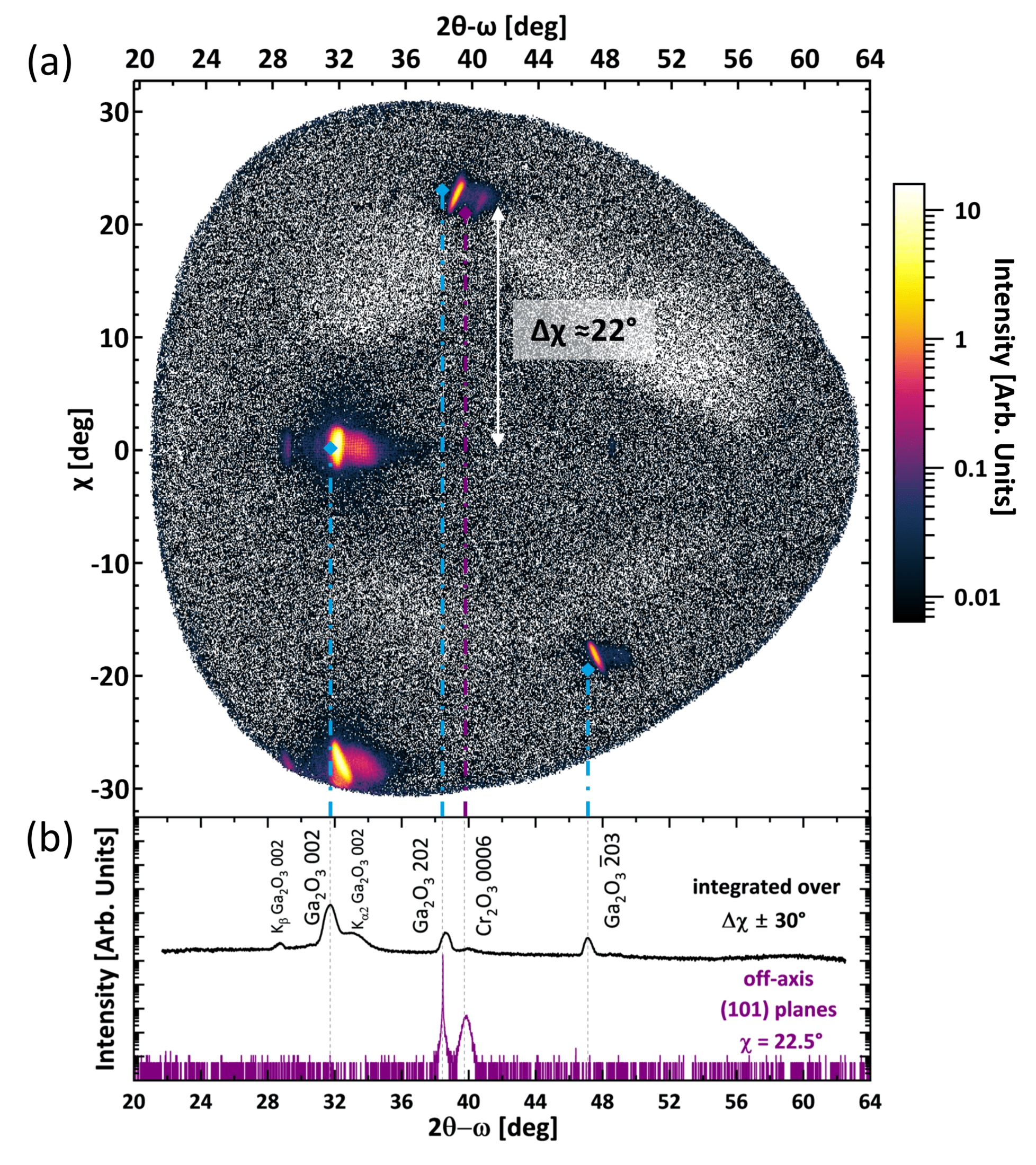}
    \caption{\label{fig:Fig2} In (a), 2D XRD intensity map obtained with powder-diffractometer for 2$\theta$-$\omega$ scans acquired in a variable range of $\Delta\chi$ = $\pm$ 30$^\circ$ with respect to the out-of-plane orientation. In (b), reported in black, the resulting 2$\theta$-$\omega$ scan obtained by integrating the intensity over the whole $\chi$ range. Light-blue and purple dashed lines highlight substrate and epilayer diffraction peaks, respectively. In (b), reported in purple, the symmetric 2$\theta$-$\omega$ scan acquired with the HR-XRD diffractometer at fixed $\chi$= 22.5$^\circ$ to probe off-axes (101) Ga$_2$O$_3$ planes. The slight shift in the Ga$_2$O$_3$ 202 diffraction peak between the HR-XRD and powder-XRD measurements arises from the fact that substrate peak alignment is not performed prior to powder-XRD measurement.}
\end{figure}

The Cr$_2$O$_3$ epilayers grown on (001) and ($\bar{2}$01) Ga$_2$O$_3$ are initially characterized following the epitaxial relationship reported in literature, \textit{i.e.}, Cr$_2$O$_3$ (0001) || Ga$_2$O$_3$ ($\bar{2}$01), for both out-of-plane substrate orientations.\cite{callahan_reliable_2024, ghosh_epitaxial_2019} As reported in Figure \ref{fig:Fig1} (a), the out-of-plane epitaxial relationship for Cr$_2$O$_3$/($\bar{2}$01) Ga$_2$O$_3$ is easily confirmed from an on-axis ($\chi = 0^\circ$) symmetric 2$\theta$-$\omega$ scan, where both substrate ($\bar{2}$01) and epilayer (0001) planes are probed. On the contrary, the characterization of Cr$_2$O$_3$/(001) Ga$_2$O$_3$ is less trivial. From an on-axis symmetric 2$\theta$-$\omega$ scan only the (001) planes of  Ga$_2$O$_3$ are detected (see SI - Figure S1(b)). This indicates that no Cr$_2$O$_3$ planes are oriented parallel to the substrate surface. Taking the results from the ($\bar{2}$01) substrate, we measured along the Ga$_2$O$_3$ [$\bar{2}$01] direction, accessed by tilting $\chi$ at 50$^\circ$ in-plane, and were still only able to observe Ga$_2$O$_3$ peaks. This result is contradictory with previous results,\cite{callahan_reliable_2024} likely due to discrepancies with alignment procedures and the use of lower-resolution receiving optics.\\
We found that the Cr$_2$O$_3$ could only be observed when aligning to the Ga$_2$O$_3$ (101) planes, which are inclined from the surface by $\chi$ $\simeq$ 22.5$^\circ$, as shown in Figure~\ref{fig:Fig1}(b). The identification of this specific substrate planes was only possible through a complementary investigation we performed, using an additional powder diffractometer, that allows to probe the sample on a wider range of $\chi$ angles, \textit{i.e.}, access out-of-plane and multiple in-plane diffractions simultaneously. Figure~\ref{fig:Fig2}(a) reports the 2D XRD intensity map acquired on a $\Delta\chi$ = $\pm30^\circ$, around the out-of-plane 002 diffraction peak (2$\theta$ = 31.73$^\circ$ and $\chi = 0^\circ$). In Figure~\ref{fig:Fig2}(b), the black $2\theta$-$\omega$ spectrum is obtained by integrating the intensity over the entire $\chi$ range. In addition to the substrate reflections, an additional diffraction feature (purple asterisk) was detected at an off-axis angle of $\Delta\chi \approx 22^\circ$ with respect to the out-of-plane direction, and at 2$\theta \approx 40^\circ$. This is assigned to the Cr$_2$O$_3$ (0006) plane: bulk Cr$_2$O$_3$ (0006) 2$\theta = 39.7482^\circ$. The close proximity in $\chi$ of the Cr$_2$O$_3$ 0006 diffraction peak with a strong substrate reflection, located at $2\theta = 38.4^\circ$, is suggestive of epitaxial relationship. By cross-checking the possible substrate reflections around this $2\theta$ position, the 202 and $\bar{4}02$ are found at very similar angular positions, \textit{i.e.}, $\Delta 2\theta = 0.0111^\circ$. This very small angular separation can therefore explain the possible misassignment of the epitaxial relationship reported previously.\cite{callahan_reliable_2024} To unambiguously identify the detected diffraction peak, the interplanar angles between the $(202)$, $(\bar{4}02)$, and the out-of-plane (001) are evaluated. The Ga$_2$O$_3$ $(\bar{2}01)$ planes are located at an off-axis angle of $\chi = 50.1^\circ$, whereas the $(202)$ planes are expected at $\chi = 22.5^\circ$. The latter value is consistent with the position of the diffraction feature observed in the 2D XRD map (Figure \ref{fig:Fig2}(a)). This agreement allows us to confidently assign the observed peak to the Ga$_2$O$_3$ 202 reflection and to establish the following epitaxial relationship: {Cr$_2$O$_3$}\,(0001) || {Ga$_2$O$_3$}\,(101) on (001) out-of-plane Ga$_2$O$_3$. To further confirm this evidence, a dedicated HR-XRD off-axis symmetric 2$\theta$-$\omega$ scan is performed, targeting specifically the (101) Ga$_2$O$_3$ planes. As reported in Figure \ref{fig:Fig1}(b) and \ref{fig:Fig2}(b), both epilayer and substrate peaks are visible with also appearance of higher order reflections.\\

\begin{figure}[h]
\centering
    \includegraphics[width=\columnwidth]{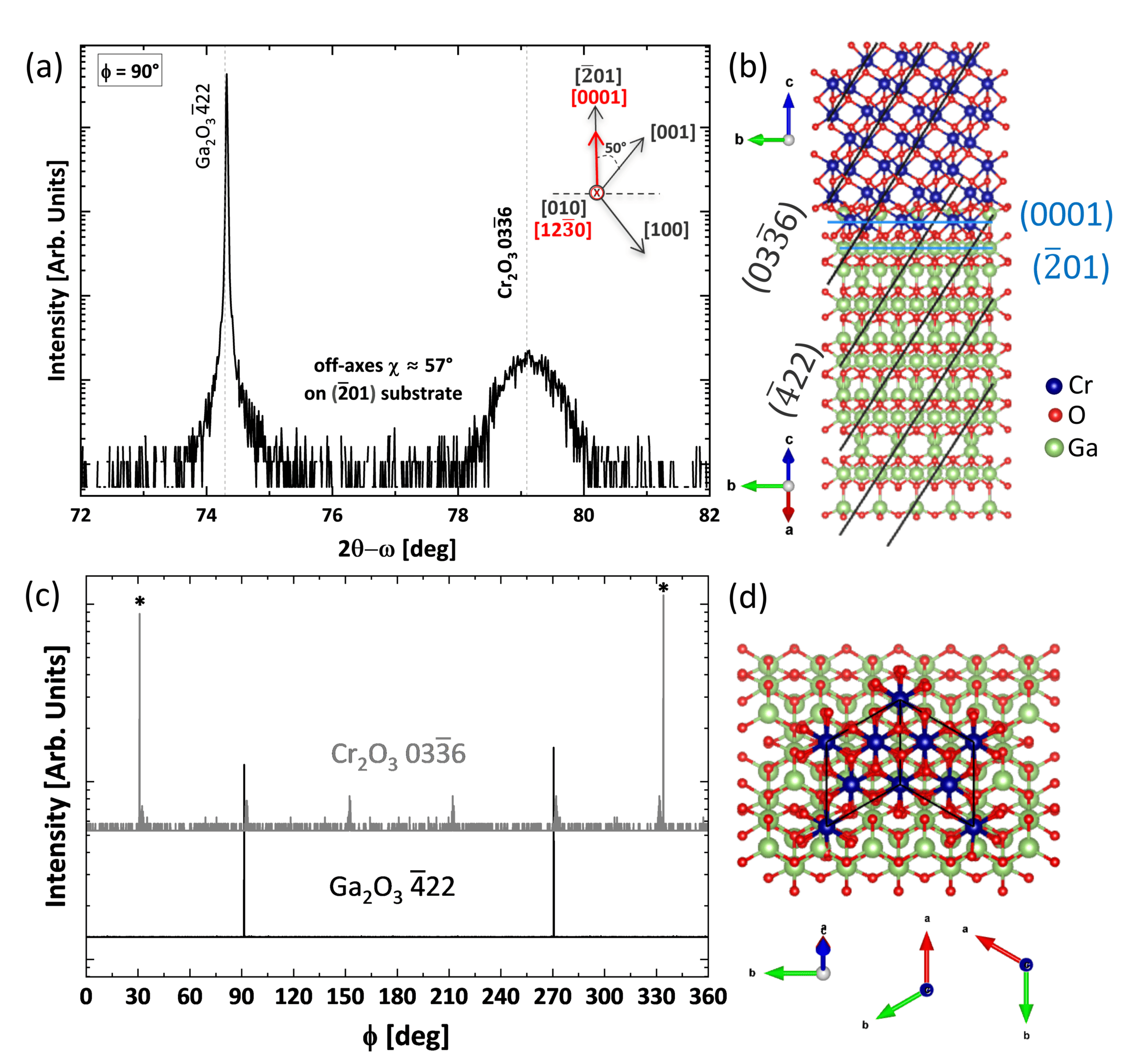}
    \caption{\label{fig:Fig3}In (a), symmetric 2$\theta$-$\omega$ off-axis ($\chi = 57^\circ$) scan of ($\bar{4}$22) Ga$_2$O$_3$ and (03$\bar{3}$6) Cr$_2$O$_3$ for a Cr$_2$O$_3$/($\bar{2}$01)Ga$_2$O$_3$ sample. In (b), schematic representation of the epilayer/substrate stack with ($\bar{4}$22), (03$\bar{3}$6), ($\bar{2}01)$ and (0001) planes reported. The in-plane vector alignment required for these planes to be parallel is: Cr$_2$O$_3$ $[12\bar{3}0]$ $\parallel$ Ga$_2$O$_3$ $[010]$. In panel (a) we report also the visualization in cross-section of the vectors alignment in-plane and out-of-plane between Ga$_2$O$_3$ (in black) and Cr$_2$O$_3$ (in red). This representation is rotated of 90$^\circ$ around the vertical z axis with respect to the representation reported in (b). In (c), $\phi$ scans for the epilayer (03$\bar{3}$6) and substrate ($\bar{4}$22), in grey and black, respectively. The two most intense peaks on the (03$\bar{3}$6) Cr$_2$O$_3$ phi scan, labeled with black asterisks, correspond to ($\bar{9}10$) substrate planes (expected at $\chi = 57.1^\circ$, 2$\theta = 79^\circ$, satisfying Bragg diffraction conditions in the reported $\phi$ scan). The six fold symmetry recorded for the epilayer is indicative of the presence of two in-plane 60$^\circ$-rotated twin domains, as represented in (d), from top-view.}
\end{figure}

For the Cr$_2$O$_3$/$(\bar{2}01)$ Ga$_2$O$_3$ sample, although the out-of-plane orientation is well established (see Figure \ref{fig:Fig1}(a)), the in-plane epitaxial relationship required further investigation. Here, as shown in Figure~\ref{fig:Fig3}(a), the Cr$_2$O$_3$ $03\bar{3}6$ and Ga$_2$O$_3$ $\bar{4}42$ reflections were identified in an off-axis ($\chi \approx 57^\circ$) symmetric $2\theta$-$\omega$ scan. To satisfy this experimentally observed off-axis crystallographic alignment as well the reported out-of-plane, \textit{i.e.}, (0001)  Cr$_2$O$_3$ || ($\bar{2}$01) Ga$_2$O$_3$, the in-plane epitaxial relationship must be as follows: Cr$_2$O$_3$ $[12\bar{3}0]$ $\parallel$ Ga$_2$O$_3$ $[010]$. A schematic representation of the resulting epitaxial alignment is provided in Figure~\ref{fig:Fig3}(b) in cross sectional view, with the investigated off- and on-axis planes of epilayer and substrate reported, to highlight their parallel alignment. To further facilitate the visualization of this epitaxial relationship, we built a vector representation, reported in panel (a) of Figure \ref{fig:Fig3}, for Ga$_2$O$_3$ (in black) and Cr$_2$O$_3$ (in red). In Figure~\ref{fig:Fig3}(c), we also report the corresponding $\phi$ scans for the substrate and epilayer off-axis reflections. The six diffraction peaks observed for the Cr$_2$O$_3$ $03\bar{3}6$ reflection indicate the presence of two in-plane twin domains rotated by $60^\circ$ with respect to each other. Such rotational domains have been reported in Cr$_2$O$_3$-based heterostructures and other material systems. Their presence can modify the effective in-plane transport and introduce anisotropic or domain-dependent electrical behavior, and should therefore be carefully considered when designing and interpreting devices based on these heterostructures.\cite{gao_process_2017,punugupati_strain_2014, https://doi.org/10.1002/adfm.202207821} The in-plane alignment and the rotational domain presence is represented in top-view in Figure \ref{fig:Fig3}(d).

The in-plane epitaxial relationship of Cr$_2$O$_3$/(001) Ga$_2$O$_3$ has not be determined unambiguously. Nevertheless, the experimentally established relationship, Cr$_2$O$_3$ (0001) $\parallel$ Ga$_2$O$_3$ (101), provides clear evidence of epitaxial relationship of the Cr$_2$O$_3$ epilayer. While this experimental result is not sufficient to fully define the in-plane alignment, they represent a key step toward establishing the complete epitaxial relationship. Further structural characterization is therefore required to resolve the in-plane orientation unambiguously.\\

Given the successful characterization of the Cr$_2$O$_3$/(001) Ga$_2$O$_3$ system, achieved here for the first time, and motivated by the challenges reported in literature for the characterization of the NiO/(001) Ga$_2$O$_3$ interface,\cite{nakagomi_crystal_2020} we applied the same crystallographic characterization strategy to NiO/Ga$_2$O$_3$ heterostructures. As detected for Cr$_2$O$_3$ (0001) planes, we observed that the NiO (111) planes are parallel to either the (101) or $(\bar{2}01)$ Ga$_2$O$_3$ planes, depending on the out-of-plane substrate orientation. The corresponding on- and off-axis XRD $2\theta$-$\omega$ scans are reported in the Figure \ref{fig:Fig1} (b,c). Furthermore, we performed a detailed investigation of the crystallographic structure of the NiO/(001) Ga$_2$O$_3$ interface by systematically examining the (101), $(\bar{2}01)$, and $(\bar{4}01)$ Ga$_2$O$_3$ planes (Figure S2(a)). This analysis provided additional crystallographic constraints on the NiO orientation and enabled us to construct the corresponding vector alignment (Figure~S2(b)). Importantly, these additional plane alignments provide a more complete crystallographic framework for identifying and characterizing the NiO epitaxial orientation on the low-symmetry (001) Ga$_2$O$_3$ surface.\\

Combining the results obtained so far, two distinct orientations of the Cr$_2$O$_3$ (0001) and NiO (111) planes are identified, depending on the nominal out-of-plane orientation of Ga$_2$O$_3$. Specifically for (001) Ga$_2$O$_3$: Cr$_2$O$_3$ (0001) and NiO (111) $\parallel$ Ga$_2$O$_3$ (101). For ($\bar{2}$01) Ga$_2$O$_3$: Cr$_2$O$_3$ (0001) and NiO (111) $\parallel$ Ga$_2$O$_3$ ($\bar{2}$01) (see the comparison of 2$\theta$-$\omega$ scans in SI - Figure S3 and S4). A schematic representation of the Cr$_2$O$_3$/Ga$_2$O$_3$ and NiO/Ga$_2$O$_3$ heterostructures are shown in Table~\ref{tab:epitaxial_relationships}, clearly illustrating the alignment with the Ga$_2$O$_3$ ($\bar{2}$01) and (101), respectively. The gray boxes overlaid at the interface serves to emphasize that the schematic is intended to provide a straightforward visualization of the macroscopic epitaxial alignment between the epilayer and substrate. The exact atomic-scale interfacial arrangement, however, requires more detailed experimental and computational investigation, such as the combined TEM and atomistic modeling approach recently reported for the NiO/Ga$_2$O$_3$ interface.\cite{smeaton2026revealingatomicstructurenioga2o3} This visualization enables also a direct comparison between the two \textit{p}-type oxides, highlighting their remarkably similar growth behavior on Ga$_2$O$_3$ despite their distinct crystal structures. The reported tilted epitaxial alignment of the Cr$_2$O$_3$ $(0001)$ and NiO $(111)$ planes can be explained by considering the matching of their oxygen sublattices with that of Ga$_2$O$_3$. In particular, the $(\bar{2}01)$ and $(101)$ Ga$_2$O$_3$ planes exhibit equivalent oxygen sublattices (see SI, Figure~S5), providing equivalent structural templates matching Cr$_2$O$_3$ $(0001)$ and NiO $(111)$ planes. The equivalence of these two Ga$_2$O$_3$ planes was also recently reported by Oshima et al.,\cite{oshima_mapping_2026} based on a description of the monoclinic Ga$_2$O$_3$ structure in terms of its equivalent cubic oxygen sublattice. The preferential alignment of the epilayers with the $(101)$ Ga$_2$O$_3$ plane observed for the $(001)$-oriented substrate may therefore be associated with its closer proximity to the substrate normal compared with the $(\bar{2}01)$ plane, together with the lower strain of the oxygen sublattices reported for this configuration.\\

\begin{table}[t]
\centering
\caption{Schematic representation, generated using VESTA, of the epitaxial relationships for Cr$_2$O$_3$ (top panel) and NiO (bottom panel) grown on ($\bar{2}$01) and (001) oriented Ga$_2$O$_3$ substrates. The alignments of the Cr$_2$O$_3$ (0001) and NiO (111) planes, depending on the substrate orientation, is highlighted.}
\label{tab:epitaxial_relationships}

\begin{tabular}{c c c}
\hline \hline
&
\textbf{$(\bar{2}01)$ Ga$_2$O$_3$} &
\textbf{$(001)$ Ga$_2$O$_3$} \\
\hline \hline

\raisebox{8\height}{\textbf{Cr$_2$O$_3$}} &
\includegraphics[width=0.2\textwidth]{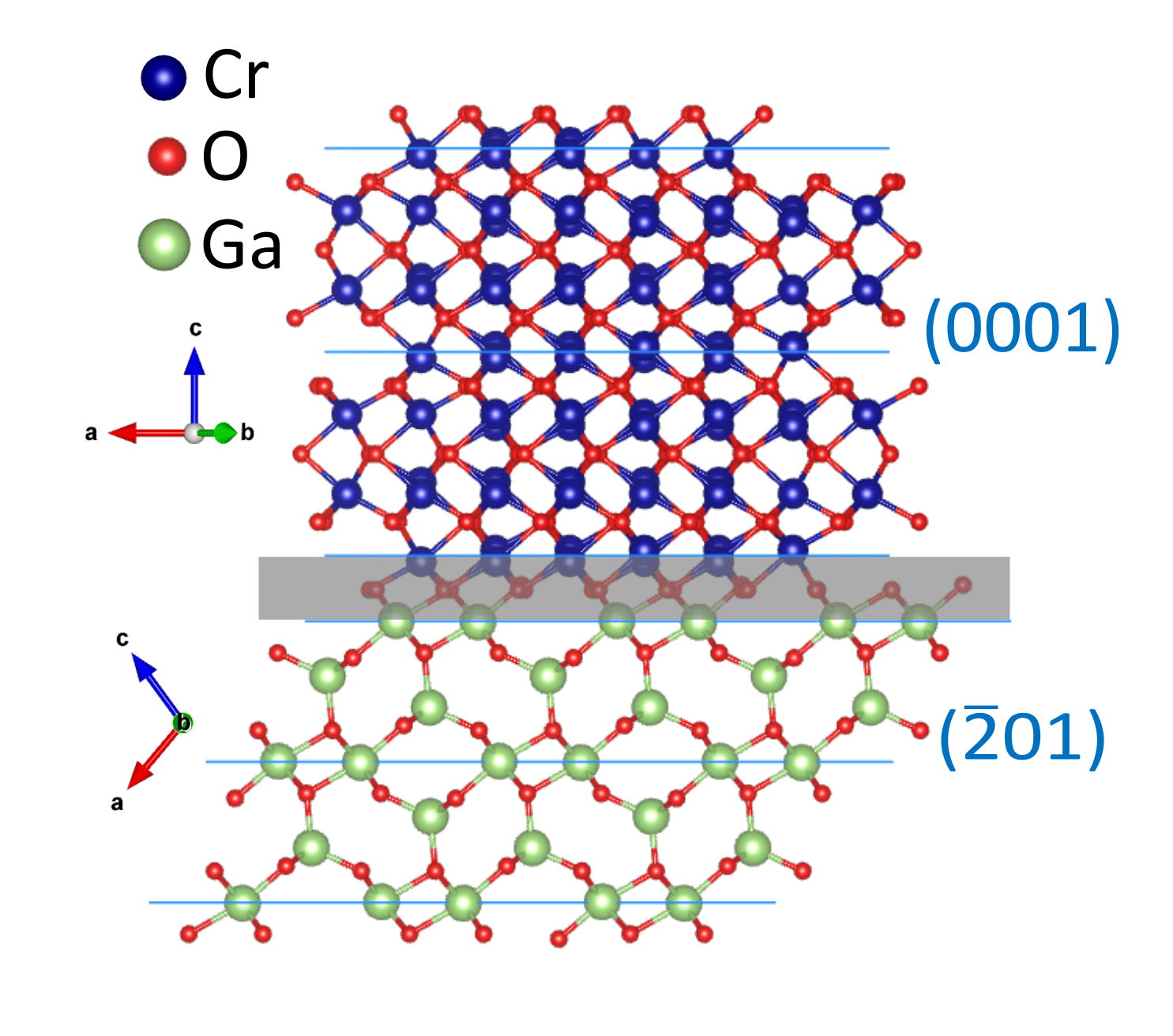} &
\includegraphics[width=0.2\textwidth]{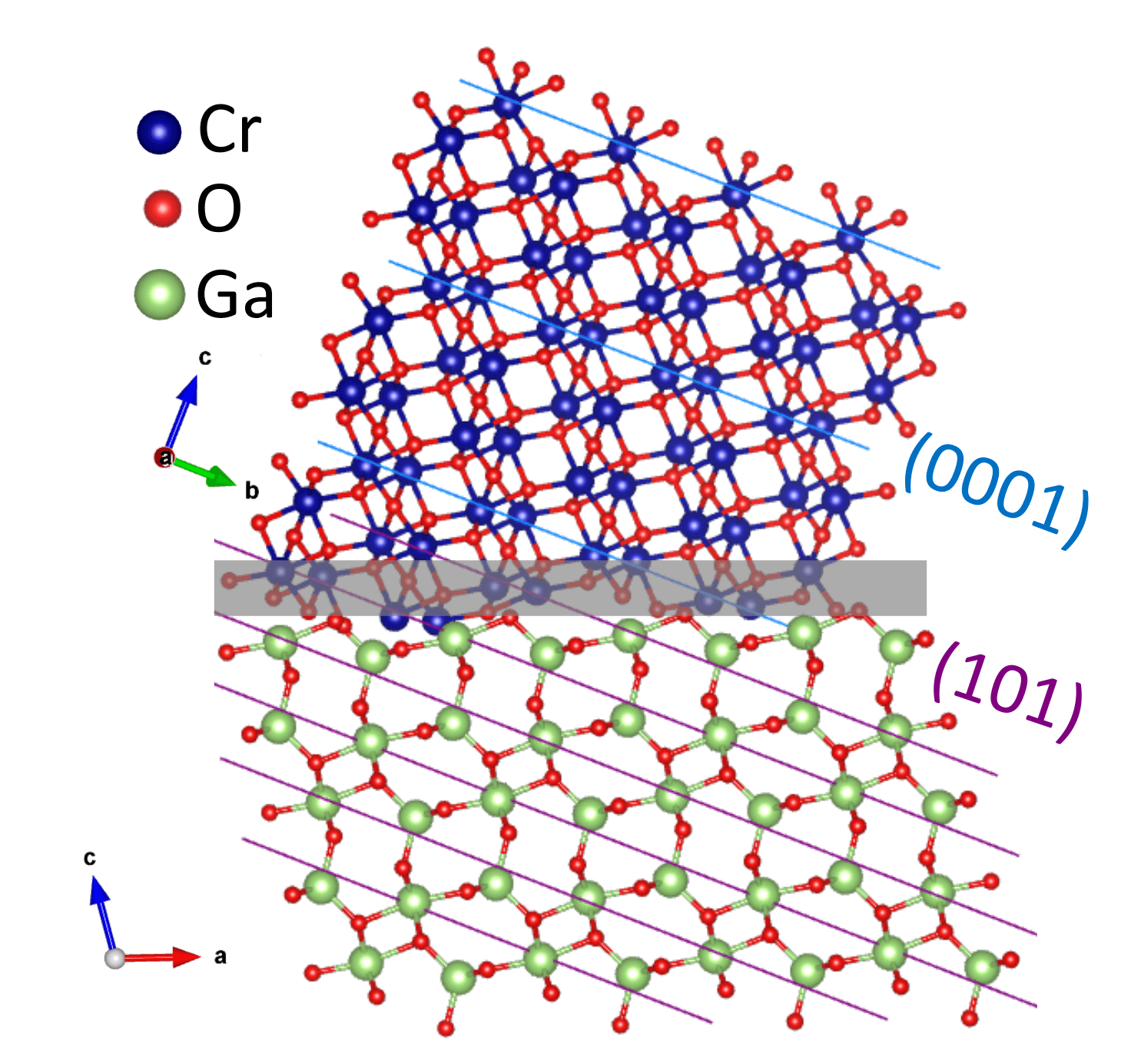}
\\[3mm]
\hline

\raisebox{8\height}{\textbf{NiO}} &
\includegraphics[width=0.2\textwidth]{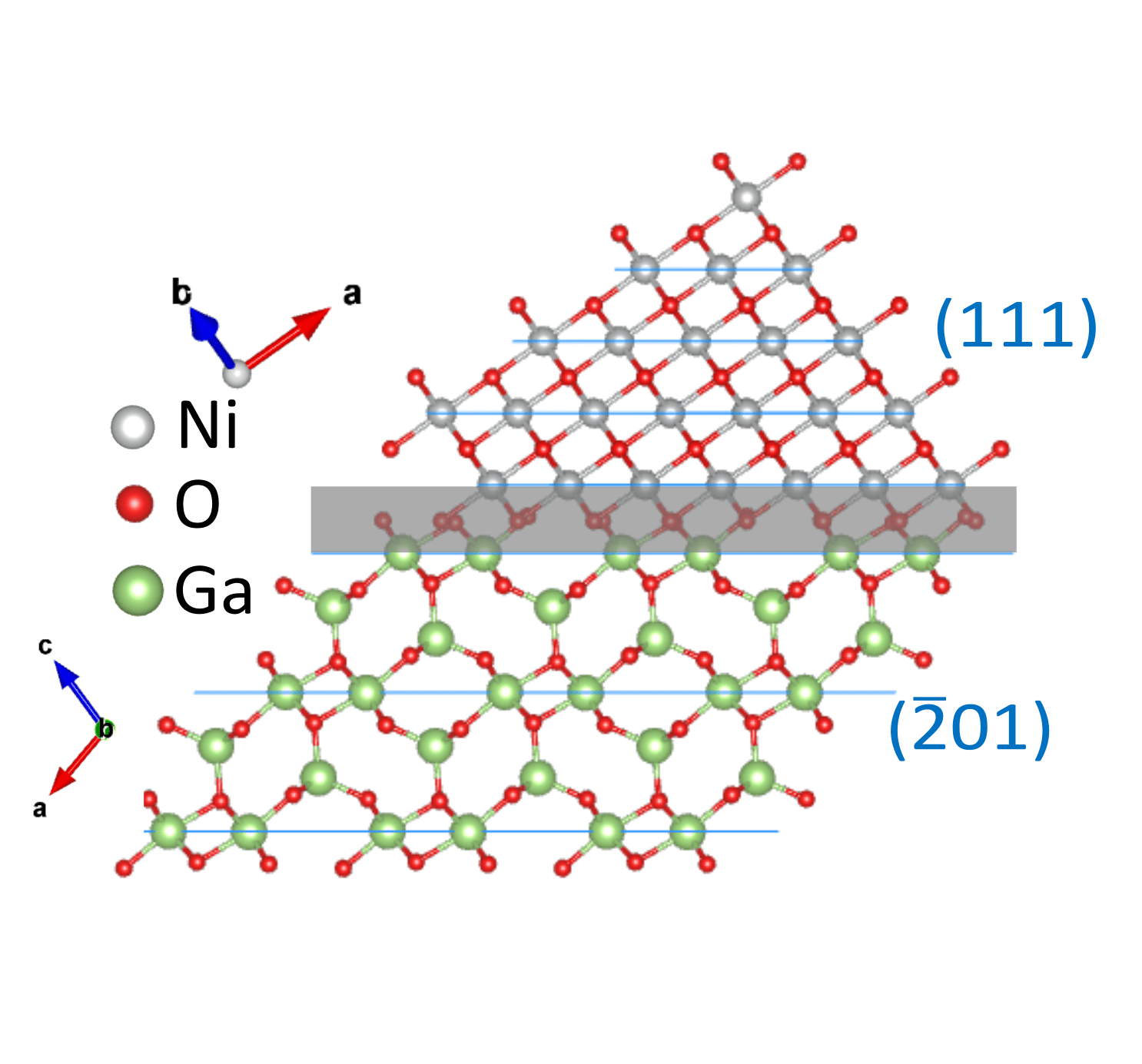} &
\includegraphics[width=0.2\textwidth]{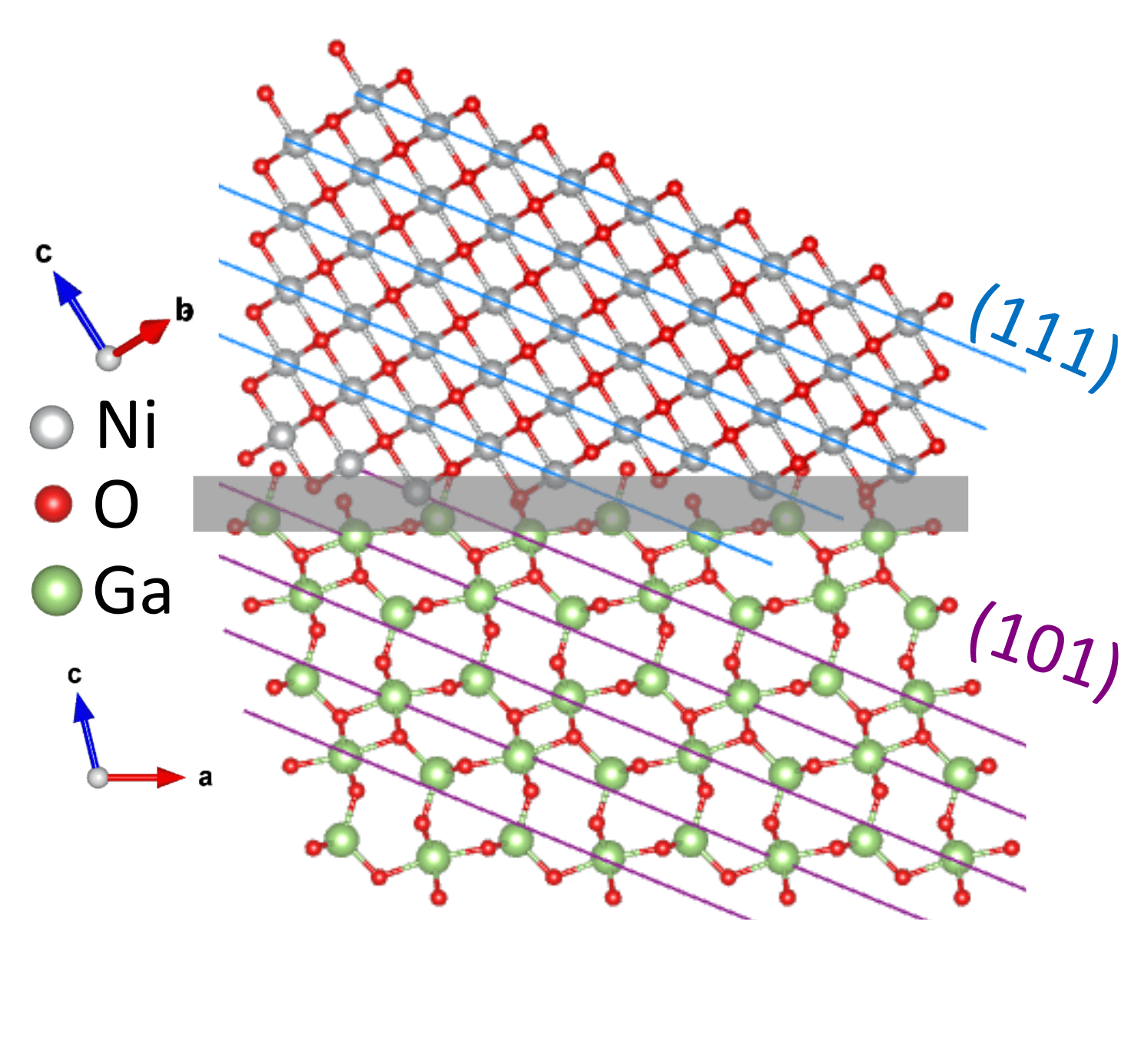}
\\
\hline \hline

\end{tabular}

\end{table}

To support the experimental findings, interface modeling was performed for the Cr$_2$O$_3$/Ga$_2$O$_3$ system to identify the most favorable interfacial atomic configurations. The detailed process of interface construction and surface-cutting procedures followed the methodology implemented in our previous work. \cite{smeaton2026revealingatomicstructurenioga2o3, Therrien_JCP_2020, stevanovic_APL_2014} The substrate and epilayer crystal lattices are generated by applying the calculated lattice parameters (see Table \ref{tab:lattice_constants}). Both O-rich and O-poor surface terminations were considered for the Ga$_2$O$_3$ ($\bar{2}$01) and (001) surfaces. The chemical-potential values used to define these surface terminations are provided in the Supplementary Information (SI - Table III). Each fixed Ga$_2$O$_3$ surface termination was then systematically matched with a series of corundum Cr$_2$O$_3$ surface orientations ($-3 \leq h,k,l \leq 3$). The structure-matching algorithm performs atom-to-atom mapping between the two materials without requiring prior knowledge of their interfacial periodicity. Following our established methodology,\cite{Therrien_JCP_2020,Therrien_PRAppl_2021} Ga$_2$O$_3$ was treated as a rigid substrate, while the Cr$_2$O$_3$ overlayer was allowed to accommodate the lattice mismatch through structural distortion.

\begin{table}[!h]
    \centering 
    \caption{Lattice parameters of bulk Ga$_2$O$_3$ and Cr$_2$O$_3$ computed from DFT.}
    \label{tab:lattice_constants}
    \renewcommand{\arraystretch}{1.2} 
    \begin{tabular}{lcccccc}
        \hline\hline 
        \multirow{2}{*}{Compound} & \multicolumn{6}{c}{Lattice Parameters (GGA)} \\ \cline{2-5} \cline{6-7}
        & $a$ [\AA] & $b$ [\AA] & $c$ [\AA] & $\alpha$ [$^\circ$] & $\beta$ [$^\circ$] & $\gamma$ [$^\circ$]\\
        \hline
        $\beta$-Ga$_2$O$_3$ & 12.43 & 3.08 & 5.88  & 90 & 103.71 & 90  \\
        $\alpha$-Cr$_2$O$_3$ & 5.05  & 5.05 & 13.83 & 90 & 90     & 120 \\
        \hline\hline 
    \end{tabular}
\end{table}

\begin{figure}
\centering
    \includegraphics[width=\columnwidth]{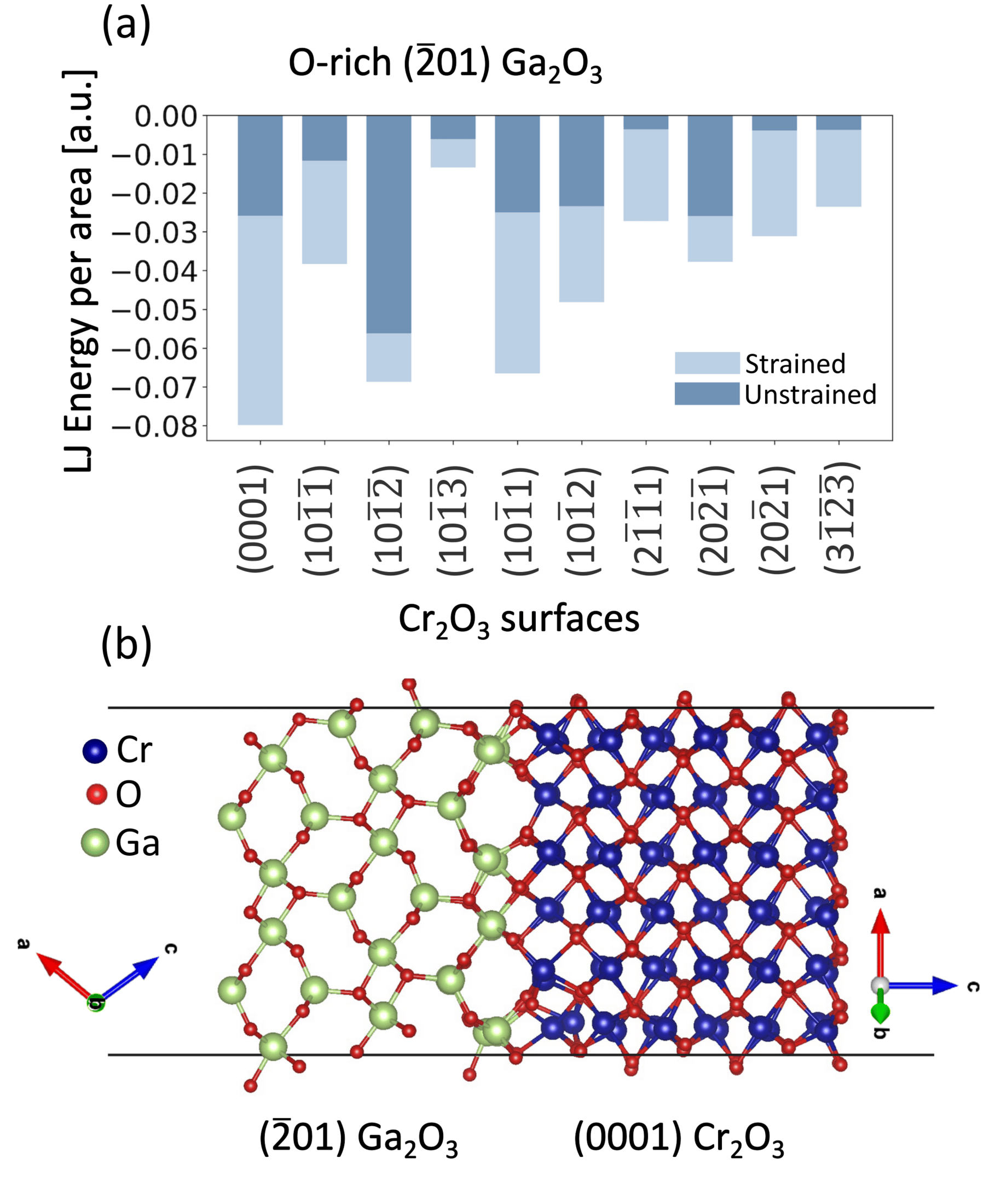}
    \caption{\label{fig:Fig4} Structure matching results for the Cr$_{2}$O$_{3}$/($\bar{2}$01) Ga$_{2}$O$_{3}$ system. In (a) the evaluated Lennard-Jones (LJ) energy per unit area for O-rich terminated Ga$_{2}$O$_{3}$($\bar{2}$01)/Cr$_{2}$O$_{3}$ ($hkl$) interfaces with $-3 \leq h,k,l \leq 3$ and a maximum allowable area strain of 8\%. A lower LJ energy per unit area indicates better interface matching. Light and dark blue bars represent the goodness of match with and without strain, respectively. (b) Atomic model of the most favorable semi-coherent interface, formed between the O-rich Ga$_{2}$O$_{3}$($\bar{2}$01) termination and the Cr$_{2}$O$_{3}$ (0001) surface.}
\end{figure}

For the Ga$_2$O$_3$ ($\bar{2}$01) surface, the structural search indicates that the O-rich termination in combination with the Cr$_2$O$_3$ (0001) surface provides the most favorable configuration when the strain associated with epitaxial nucleation is taken into account. As shown in Figure \ref{fig:Fig4}(a), the Cr$_2$O$_3$ (0001) interface exhibits the lowest Lennard-Jones (LJ) energy per area among the strained configurations (light-blue bars). Even though the Cr$_2$O$_3$ (10$\bar{1}$$\bar{2}$) surface yields the absolute lowest LJ energy for the corresponding unstrained configuration (dark-blue bars), the strained configuration is expected to more closely represent the physical conditions during epitaxial nucleation, where the growing epilayer initially accommodates the substrate lattice rather than adopting its fully relaxed bulk structure. Therefore, the modeling results support the Cr$_2$O$_3$ (0001) || Ga$_2$O$_3$ ($\bar{2}$01) orientation observed experimentally. In SI - Figure S6, we report the calculations relative to the O-poor terminated ($\bar{2}$01) Ga$_2$O$_3$/Cr$_2$O$_3$ interfaces, resulting in overall higher LJ energy. For the (001)-oriented Ga$_2$O$_3$ substrate, as observed experimentally, identifying the most energetically favorable Cr$_2$O$_3$ surface match is less straightforward. The absence of Cr$_2$O$_3$ planes oriented parallel to the Ga$_2$O$_3$ (001) surface prevents the straightforward identification of a physically reasonable interface configuration. Considering higher-index Cr$_2$O$_3$ planes could provide additional candidate configurations; however, this would substantially increase the computational cost and complexity of the interface modeling. In SI - Figure S7, we report the calculation performed for O-rich and O-poor terminated Ga$_2$O$_3$ (001)/Cr$_2$O$_3$ surface matching.

In conclusion, this work provides an in-depth analysis of the crystallographic epitaxial relationship of Cr$_2$O$_3$ and NiO/Ga$_2$O$_3$ heterostructures, for two different out-of-plane substrate orientations, \textit{i.e.}, (001) and $(\bar{2}01)$. By combining X-ray diffraction with crystal-structure visualization tool, we identified two possible alignments for Cr$_2$O$_3$ (0001) and NiO (111) planes, found to be  parallel either to $(\bar{2}01)$ or (101) Ga$_2$O$_3$, according to out-of-plane substrate orientation. We explained this tilted epitaxial relationship in terms of match between epilayer and substrate oxygen anion sublattices and substrate planes equivalence. The in-plane alignment for the Cr$_2$O$_3$/($\bar{2}$01) Ga$_2$O$_3$ was fully disclosed as: Cr$_2$O$_3$ $[12\bar{3}0]$ $\parallel$ Ga$_2$O$_3$ $[010]$ and the presence of two 60$^\circ$ rotated in-plane domains was recorded.
Furthermore, this work highlights the challenges of epilayer growth on (001) Ga$_2$O$_3$. Despite its widespread use in device applications, this orientation complicates heterointerface characterization. Experimentally, the absence of low-index epilayer planes parallel to the substrate surface makes the epitaxial relationship difficult to resolve. Computationally, it increases the complexity of identifying favorable interface configurations. Nevertheless, the methodology presented here provides a practical framework for characterizing heterostructures on (001) Ga$_2$O$_3$ and advancing the understanding of these technologically relevant interfaces.

\begin{acknowledgments}
This work was supported as part of the A Center for Power Electronics Materials and Manufacturing Exploration (APEX) Energy Frontier Research Center funded by the U.S. DOE, Office of Science, Basic Energy Sciences. This work was authored in part by the National Laboratory of the Rockies (NLR) for the U.S. Department of Energy (DOE) under Contract No. DE-AC36-08GO28308. The views expressed in the article do not necessarily represent the views of the DOE or the U.S. Government
\end{acknowledgments}

\section{References}
\bibliography{Reference}

@article{https://doi.org/10.1002/adfm.202207821,
author = {Mazzolini, Piero and Fogarassy, Zsolt and Parisini, Antonella and Mezzadri, Francesco and Diercks, David and Bosi, Matteo and Seravalli, Luca and Sacchi, Anna and Spaggiari, Giulia and Bersani, Danilo and Bierwagen, Oliver and Janzen, Benjamin Moritz and Marggraf, Marcella Naomi and Wagner, Markus R. and Cora, Ildiko and Pécz, Béla and Tahraoui, Abbes and Bosio, Alessio and Borelli, Carmine and Leone, Stefano and Fornari, Roberto},
title = {Silane-Mediated Expansion of Domains in Si-Doped $\kappa$-Ga2O3 Epitaxy and its Impact on the In-Plane Electronic Conduction},
journal = {Advanced Functional Materials},
volume = {33},
number = {2},
pages = {2207821},
doi = {https://doi.org/10.1002/adfm.202207821},
url = {https://advanced.onlinelibrary.wiley.com/doi/abs/10.1002/adfm.202207821},
eprint = {https://advanced.onlinelibrary.wiley.com/doi/pdf/10.1002/adfm.202207821},
year = {2023}
}

@article{li_breakdown_2024,
	title = {Breakdown up to 13.5 {kV} in {NiO}/$\beta$-{Ga2O3} {Vertical} {Heterojunction} {Rectifiers}},
	volume = {13},
	issn = {2162-8777},
	url = {https://doi.org/10.1149/2162-8777/ad3457},
	doi = {10.1149/2162-8777/ad3457},
	number = {3},
	urldate = {2026-09-15},
	journal = {ECS J. Solid State Sci. Technol.},
	author = {Li, Jian-Sian and Wan, Hsiao-Hsuan and Chiang, Chao-Ching and Yoo, Timothy Jinsoo and Yu, Meng-Hsun and Ren, Fan and Kim, Honggyu and Liao, Yu-Te and Pearton, Stephen J.},
	month = mar,
	year = {2024},
	note = {Publisher: IOP Publishing},
	pages = {035003},
}

@article{nishinaka_microstructures_2018,
	title = {Microstructures and rotational domains in orthorhombic $\varepsilon$-{Ga}$_{\textrm{2}}$ {O}$_{\textrm{3}}$ thin films},
	volume = {57},
	issn = {0021-4922, 1347-4065},
	url = {https://iopscience.iop.org/article/10.7567/JJAP.57.115601},
	doi = {10.7567/JJAP.57.115601},
	number = {11},
	urldate = {2026-09-15},
	journal = {Jpn. J. Appl. Phys.},
	author = {Nishinaka, Hiroyuki and Komai, Hiroki and Tahara, Daisuke and Arata, Yuta and Yoshimoto, Masahiro},
	month = nov,
	year = {2018},
	pages = {115601},
}

@article{park_rhombohedral_2008,
	title = {Rhombohedral epitaxy of cubic {SiGe} on trigonal c-plane sapphire},
	volume = {310},
	issn = {0022-0248},
	url = {https://www.sciencedirect.com/science/article/pii/S0022024808001371},
	doi = {https://doi.org/10.1016/j.jcrysgro.2008.02.010},
	number = {11},
	journal = {Journal of Crystal Growth},
	author = {Park, Yeonjoon and King, Glen C. and Choi, Sang H.},
	year = {2008},
	pages = {2724--2731},
}

@article{https://doi.org/10.1002/pssb.200303368,
author = {Paskova, T. and Darakchieva, V. and Valcheva, E. and Paskov, P. P. and Monemar, B. and Heuken, M.},
title = {Growth of GaN on a-plane sapphire: in-plane epitaxial relationships and lattice parameters},
journal = {physica status solidi (b)},
volume = {240},
number = {2},
pages = {318-321},
doi = {https://doi.org/10.1002/pssb.200303368},
url = {https://onlinelibrary.wiley.com/doi/abs/10.1002/pssb.200303368},
eprint = {https://onlinelibrary.wiley.com/doi/pdf/10.1002/pssb.200303368},
year = {2003}
}

@article{nakagomi_crystal_2020,
	title = {Crystal {Orientation} of {Cubic} {NiO} {Thin} {Films} {Formed} on {Monoclinic} $\beta$-{Ga2O3} {Substrates}},
	volume = {257},
	copyright = {© 2019 WILEY-VCH Verlag GmbH \& Co. KGaA, Weinheim},
	issn = {1521-3951},
	url = {https://onlinelibrary.wiley.com/doi/abs/10.1002/pssb.201900669},
	doi = {10.1002/pssb.201900669},
	number = {5},
	urldate = {2025-11-06},
	journal = {physica status solidi (b)},
	author = {Nakagomi, Shinji and Yasuda, Takashi and Kokubun, Yoshihiro},
	year = {2020},
	note = {\_eprint: https://onlinelibrary.wiley.com/doi/pdf/10.1002/pssb.201900669},
	pages = {1900669},
}

@article{punugupati_strain_2014,
	title = {Strain induced ferromagnetism in epitaxial {Cr2O3} thin films integrated on {Si}(001)},
	volume = {105},
	issn = {0003-6951},
	url = {https://doi.org/10.1063/1.4896975},
	doi = {10.1063/1.4896975},
	number = {13},
	urldate = {2026-08-26},
	journal = {Appl. Phys. Lett.},
	author = {Punugupati, Sandhyarani and Narayan, Jagdish and Hunte, Frank},
	month = sep,
	year = {2014},
	pages = {132401},
}

@article{gao_process_2017,
	series = {{SI}: {CRYS}\_ECCG5},
	title = {The process of growing {Cr2O3} thin films on $\alpha$-{Al2O3} substrates at low temperature by r.f. magnetron sputtering},
	volume = {457},
	issn = {0022-0248},
	url = {https://www.sciencedirect.com/science/article/pii/S0022024816304171},
	doi = {10.1016/j.jcrysgro.2016.08.009},
	urldate = {2026-08-26},
	journal = {Journal of Crystal Growth},
	author = {Gao, Yin and Leiste, Harald and Stueber, Michael and Ulrich, Sven},
	month = jan,
	year = {2017},
	pages = {158--163},
}

@article{sawada_residual_1994,
	title = {Residual electron density study of chromium sesquioxide by crystal structure and scattering factor refinement},
	volume = {29},
	issn = {0025-5408},
	url = {https://www.sciencedirect.com/science/article/pii/0025540894900191},
	doi = {10.1016/0025-5408(94)90019-1},
	number = {3},
	urldate = {2026-08-26},
	journal = {Materials Research Bulletin},
	author = {Sawada, H.},
	month = mar,
	year = {1994},
	pages = {239--245},
}

@article{geller_crystal_1960,
	title = {Crystal {Structure} of $\beta$‐{Ga2O3}},
	volume = {33},
	issn = {0021-9606},
	url = {https://doi.org/10.1063/1.1731237},
	doi = {10.1063/1.1731237},
	number = {3},
	urldate = {2026-08-26},
	journal = {J. Chem. Phys.},
	author = {Geller, S.},
	month = sep,
	year = {1960},
	pages = {676--684},
}

@article{higashiwaki_-ga2o3_2022,
	title = {$\beta$-{Ga2O3} material properties, growth technologies, and devices: a review},
	volume = {32},
	issn = {2309-4710},
	shorttitle = {$\beta$-{Ga2O3} material properties, growth technologies, and devices},
	url = {https://doi.org/10.1007/s43673-021-00033-0},
	doi = {10.1007/s43673-021-00033-0},
	number = {1},
	urldate = {2026-08-26},
	journal = {AAPPS Bull.},
	author = {Higashiwaki, Masataka},
	month = jan,
	year = {2022},
	pages = {3},
}

@article{nakagomi_crystal_2012,
	title = {Crystal orientation of $\beta$-{Ga2O3} thin films formed on c-plane and a-plane sapphire substrate},
	volume = {349},
	issn = {0022-0248},
	url = {https://www.sciencedirect.com/science/article/pii/S0022024812002722},
	doi = {https://doi.org/10.1016/j.jcrysgro.2012.04.006},
	number = {1},
	journal = {Journal of Crystal Growth},
	author = {Nakagomi, Shinji and Kokubun, Yoshihiro},
	year = {2012},
	pages = {12--18},
}

@article{oshima_mapping_2026,
	title = {Mapping primary crystallographic planes in $\beta$-{Ga2O3} based on a pseudo-cubic oxygen sublattice},
	volume = {65},
	issn = {1347-4065},
	url = {https://doi.org/10.35848/1347-4065/ae42ac},
	doi = {10.35848/1347-4065/ae42ac},
	number = {3},
	urldate = {2026-03-02},
	journal = {Jpn. J. Appl. Phys.},
	author = {Oshima, Takayoshi},
	month = feb,
	year = {2026},
	note = {Publisher: IOP Publishing},
	pages = {038003},
}

@article{kokubun_all-oxide_2016,
	title = {All-oxide p–n heterojunction diodes comprising p-type {NiO} and n-type $\beta$-{Ga2O3}},
	volume = {9},
	issn = {1882-0786},
	url = {https://iopscience.iop.org/article/10.7567/APEX.9.091101/meta},
	doi = {10.7567/APEX.9.091101},
	number = {9},
	urldate = {2025-06-24},
	journal = {Appl. Phys. Express},
	author = {Kokubun, Yoshihiro and Kubo, Shohei and Nakagomi, Shinji},
	month = aug,
	year = {2016},
	note = {Publisher: IOP Publishing},
	pages = {091101},
}

@article{momma_vesta3_2011,
  author  = {Momma, Koichi and Izumi, Fujio},
  title   = {VESTA3 for three-dimensional visualization of crystal, volumetric and morphology data},
  journal = {Journal of Applied Crystallography},
  volume  = {44},
  number  = {6},
  pages   = {1272--1276},
  year    = {2011},
  doi     = {10.1107/S0021889811038970}
}

@misc{smeaton2026revealingatomicstructurenioga2o3,
      title={Revealing the Atomic Structure of NiO/Ga$_{2}$O$_{3}$ Interfaces}, 
      author={Michelle A. Smeaton and Krishna Acharya and Anna Sacchi and Renae N. Gannon and M. Brooks Tellekamp and Andriy Zakutayev and Vladan Stevanovic and Steven R. Spurgeon},
      year={2026},
      eprint={2608.10226},
      archivePrefix={arXiv},
      primaryClass={cond-mat.mtrl-sci},
      url={https://arxiv.org/abs/2608.10226}, 
}

@misc{liu_electrical_2025,
	title = {Electrical {Stability} of {Cr2O3}/{\textbackslash}b\{eta\}-{Ga2O3} and {NiOx}/{\textbackslash}b\{eta\}-{Ga2O3} {Heterojunction} {Diodes}},
	url = {http://arxiv.org/abs/2512.11264},
	doi = {10.48550/arXiv.2512.11264},
	urldate = {2026-01-09},
	publisher = {arXiv},
	author = {Liu, Yizheng and Wang, Haochen and Peterson, Carl and Saha, Chinmoy Nath and Walle, Chris G. Van de and Krishnamoorthy, Sriram},
	month = dec,
	year = {2025},
	note = {arXiv:2512.11264 [physics]},
}

@article{ghosh_epitaxial_2019,
	title = {Epitaxial growth and interface band alignment studies of all oxide $\alpha$-{Cr2O3}/$\beta$-{Ga2O3} p-n heterojunction},
	volume = {115},
	issn = {0003-6951},
	url = {https://doi.org/10.1063/1.5100589},
	doi = {10.1063/1.5100589},
	number = {6},
	urldate = {2025-12-02},
	journal = {Appl. Phys. Lett.},
	author = {Ghosh, Sahadeb and Baral, Madhusmita and Kamparath, Rajiv and Choudhary, R. J. and Phase, D. M. and Singh, S. D. and Ganguli, Tapas},
	month = aug,
	year = {2019},
	pages = {061602},
}

@article{ghosh_evaluation_2021,
	title = {Evaluation of valence band offset and its non-commutativity at all oxide $\alpha$-{Cr2O3}/$\beta$-{Ga2O3} heterojunction from photoelectron spectroscopy},
	volume = {130},
	issn = {0021-8979},
	url = {https://doi.org/10.1063/5.0046538},
	doi = {10.1063/5.0046538},
	number = {17},
	urldate = {2025-12-02},
	journal = {J. Appl. Phys.},
	author = {Ghosh, Sahadeb and Baral, Madhusmita and Bhattacharjee, Jayanta and Kamparath, Rajiv and Singh, S. D. and Ganguli, Tapas},
	month = nov,
	year = {2021},
	pages = {175303},
}

@misc{liu_cr2o3beta-ga2o3_2025,
	title = {{Cr2O3}/{\textbackslash}b\{eta\}-{Ga2O3} {Heterojunction} {Diodes} with {Orientation}-{Dependent} {Breakdown} {Electric} {Field} up to 12.9 {MV}/cm},
	url = {http://arxiv.org/abs/2511.20885},
	doi = {10.48550/arXiv.2511.20885},
	urldate = {2026-01-09},
	publisher = {arXiv},
	author = {Liu, Yizheng and Wang, Haochen and Peterson, Carl and Speck, James S. and Walle, Chris Van De and Krishnamoorthy, Sriram},
	month = nov,
	year = {2025},
	note = {arXiv:2511.20885 [cond-mat]},
}

@misc{egbo_epitaxial_2025,
	title = {Epitaxial growth and semiconductor properties of {NiGa2O4} spinel for {Ga2O3}/{NiO} interfaces},
	url = {http://arxiv.org/abs/2512.20841},
	doi = {10.48550/arXiv.2512.20841},
	urldate = {2026-03-04},
	publisher = {arXiv},
	author = {Egbo, Kingsley and Garrity, Emily M. and Gowda, Shivashree Shivamade and Zare, Saman and Scott, Ethan A. and Teeter, Glenn and Tellekamp, Brooks and Stevanovic, Vladan and Hopkins, Patrick E. and Zakutayev, Andriy and Haegel, Nancy},
	month = dec,
	year = {2025},
	note = {arXiv:2512.20841 [cond-mat]},
}

@article{egbo_niga2o4_2024,
	title = {{NiGa2O4} interfacial layers in {NiO}/{Ga2O3} heterojunction diodes at high temperature},
	volume = {124},
	issn = {0003-6951},
	url = {https://doi.org/10.1063/5.0194540},
	doi = {10.1063/5.0194540},
	number = {17},
	journal = {Applied Physics Letters},
	author = {Egbo, Kingsley and Garrity, Emily M. and Callahan, William A. and Chae, Chris and Lee, Cheng-Wei and Tellekamp, Brooks and Hwang, Jinwoo and Stevanovic, Vladan and Zakutayev, Andriy},
	month = apr,
	year = {2024},
	pages = {173512},
}

@article{callahan_reliable_2024,
	title = {Reliable operation of {Cr2O3}:{Mg}/$\beta$-{Ga2O3} p–n heterojunction diodes at 600 °{C}},
	volume = {124},
	issn = {0003-6951, 1077-3118},
	shorttitle = {Reliable operation of {Cr2O3}},
	url = {https://pubs.aip.org/apl/article/124/15/153504/3282486/Reliable-operation-of-Cr2O3-Mg-Ga2O3-p-n},
	doi = {10.1063/5.0185566},
	number = {15},
	urldate = {2026-08-17},
	journal = {Applied Physics Letters},
	author = {Callahan, William A. and Egbo, Kingsley and Lee, Cheng-Wei and Ginley, David and O'Hayre, Ryan and Zakutayev, Andriy},
	month = apr,
	year = {2024},
	pages = {153504},
}

@article{kresse_PRB_1999,
  title={From ultrasoft pseudopotentials to the projector augmented-wave method},
  author={Kresse, Georg and Joubert, Daniel},
  journal={Physical review b},
  volume={59},
  number={3},
  pages={1758},
  year={1999},
  publisher={APS}
}

@article{Therrien_PRAppl_2021,
  title = {Theoretical Insights for Improving the Schottky-Barrier Height at the ${\mathrm{Ga}}_{2}{\mathrm{O}}_{3}/\mathrm{Pt}$ Interface},
  author = {Therrien, F\'elix and Zakutayev, Andriy and Stevanovi\ifmmode \acute{c}\else \'{c}\fi{}, Vladan},
  journal = {Phys. Rev. Appl.},
  volume = {16},
  issue = {6},
  pages = {064064},
  numpages = {12},
  year = {2021},
  month = {Dec},
  publisher = {American Physical Society},
  doi = {10.1103/PhysRevApplied.16.064064},
  url = {https://link.aps.org/doi/10.1103/PhysRevApplied.16.064064}
}

@article{Therrien_JCP_2020,
    author = {Therrien, Félix and Graf, Peter and Stevanović, Vladan},
    title = "{Matching crystal structures atom-to-atom}",
    journal = {J. Chem. Phys.},
    volume = {152},
    number = {7},
    year = {2020},
    month = {02},
    issn = {0021-9606},
    doi = {10.1063/1.5131527},
    url = {https://doi.org/10.1063/1.5131527},
    page = {074106},
}

@software{p2ptrans_2022,
  author = {Therrien, Félix},
  title = {{p2ptrans - A Structure Matching Algorithm}},
  url = {https://github.com/ftherrien/p2ptrans},
  version = {2.1.0},
  year = {2020}
}

@article{stevanovic_APL_2014,
  title={Variations of ionization potential and electron affinity as a function of surface orientation: The case of orthorhombic SnS},
  author={Stevanovi{\'c}, Vladan and Hartman, Katy and Jaramillo, R and Ramanathan, Shriram and Buonassisi, Tonio and Graf, Peter},
  journal={Applied Physics Letters},
  volume={104},
  number={21},
  year={2014},
  publisher={AIP Publishing}
}

\end{document}


\preprint{AIP/123-QED}

\title[Supplementary Information: Revealing epitaxial relationships at Ga$_2$O$_3$ interfaces with \textit{p}-type oxides.]{Supplementary Information: Revealing epitaxial relationships at Ga$_2$O$_3$ interfaces with \textit{p}-type oxides.}

\author{Anna Sacchi}
 \email{anna.sacchi@nlr.gov}
 \affiliation{Materials Science Center, National Laboratory of the Rockies, Golden, CO, USA}
 
\author{Krishna Acharya}
 \affiliation{Metallurgical and Materials Engineering Department, Colorado School of Mines, Golden, CO, USA}
 
 \author{Michelle A. Smeaton}
  \affiliation{Materials Science Center, National Laboratory of the Rockies, Golden, CO, USA}
  
  \author{Renae N. Gannon }
  \affiliation{Materials Science Center, National Laboratory of the Rockies, Golden, CO, USA}

\author{Vladan Stevanovic}
\affiliation{Metallurgical and Materials Engineering Department, Colorado School of Mines, Golden, CO, USA}

\author{Steven R. Spurgeon}
\affiliation{Materials Science Center, National Laboratory of the Rockies, Golden, CO, USA}
\affiliation{Metallurgical and Materials Engineering Department, Colorado School of Mines, Golden, CO, USA}
\affiliation{Renewable and Sustainable Energy Institute, University of Colorado Boulder, Boulder, CO, USA}

\author{M. Brooks Tellekamp}
\affiliation{Materials Science Center, National Laboratory of the Rockies, Golden, CO, USA}

\author{Andriy Zakutayev}
\affiliation{Materials Science Center, National Laboratory of the Rockies, Golden, CO, USA}

\maketitle
\renewcommand{\thefigure}{S\arabic{figure}}

\section{Crystallographic notation and interplanar angles calculation}
The crystallographic notation adopted throughout the manuscript to describe the structural objects is summarized in Table \ref{tab:crystal}. For the substrate out-of-plane orientation, we consistently refer to the crystallographic planes parallel to the substrate surface; thus, a substrate described as $(hkl)$-oriented corresponds to the $(hkl)$ planes being parallel to the surface. For Cr$_2$O$_3$, the hexagonal Miller-Bravais notation is used, \textit{i.e.}, $(hkil)$, where $i = -(h+k)$. When reporting epitaxial relationships, we refer to either crystallographic planes, $(hkl)$, or vectors, $[hkl]$, using their lowest-index equivalents, even when the experimental evidence is obtained from higher-order reflections.\\

\begin{table}[h]
\centering
\caption{Crystallographic notation applied throughout the manuscript.}
\label{tab:crystal}
\begin{tabular}{cc}
\hline
Object & Notation\\
\hline
Plane & (hkl) \\
Scattering vector & [hkl] \\
Diffraction peak & hkl \\
\hline
\end{tabular}
\end{table}

From Equation \ref{eq:chi}, the interplanar angle $\chi$, between two different planes, \textit{i.e.}, $(hkl)$ and $(h'k'l')$, is calculated. For each unit cell geometry involved in this study, \textit{i.e.}, monoclinic Ga$_2$O$_3$, trigonal Cr$_2$O$_3$ and cubic NiO, the corresponding reciprocal lattice parameters ($\alpha^*, \beta^*,\gamma^*,a^*, b^*$ and $c^*$) and the simplified equation for interplanar spacing calculation ($d_{hkl}$) are  reported in Table \ref{tab:dhkl}.
\begin{figure*}
\centering
\begin{equation}
\cos\chi = d_{hkl}d_{h'k'l'}\Big[hh'a^{*2}+kk'b^{*2}+ll'c^{*2}
+(kl'+lk')b^{*}c^{*}\cos\alpha^{*}
+(hl'+lh')a^{*}c^{*}\cos\beta^{*}
+(hk'+kh')a^{*}b^{*}\cos\gamma^{*}\Big]
\label{eq:chi}
\end{equation}
\end{figure*}

\begin{table*}
\centering
\caption{Reciprocal lattice parameters and $d_{hkl}$ equations for the monoclinic, trigonal, and cubic crystal systems.}
\label{tab:dhkl}
\renewcommand{\arraystretch}{2.0}
\begin{tabular}{@{}l c c c@{}}
\toprule
Reciprocal lattice & Monoclinic & Trigonal & Cubic \\
\midrule
$\alpha^{*}$ & $\alpha$ & $\alpha$ & $\alpha$ \\
$\beta^{*}$  & $180^{\circ}-\beta$ & $\beta$ & $\beta$ \\
$\gamma^{*}$ & $\gamma$ & $180^{\circ}-\gamma$ & $\gamma$ \\
\addlinespace
$a^{*}$ & $\dfrac{1}{a\sin\beta}$ & $\dfrac{2}{a\sqrt{3}}$ & $\dfrac{1}{a}$ \\[8pt]
$b^{*}$ & $\dfrac{1}{b}$ & $=a^{*}$ & $=a^{*}$ \\[8pt]
$c^{*}$ & $\dfrac{1}{c\sin\beta}$ & $\dfrac{1}{c}$ & $=a^{*}$ \\[8pt]
\midrule
$d_{hkl}^{2}$ &
$\dfrac{1}{h^{2}a^{*2}+k^{2}b^{*2}+l^{2}c^{*2}+2lhc^{*}a^{*}\cos\beta^{*}}$ &
$\dfrac{1}{(h^{2}+k^{2}+hk)a^{*2}+l^{2}c^{*2}}$ &
$\dfrac{1}{(h^{2}+k^{2}+l^{2})a^{*2}}$ \\[10pt]
\bottomrule
\end{tabular}
\end{table*}

\section{Gallium oxide (101) and (\textbf{$\bar{2}01$}) planes equivalence}
\begin{figure*}
    \centering
    \includegraphics[width=\textwidth]{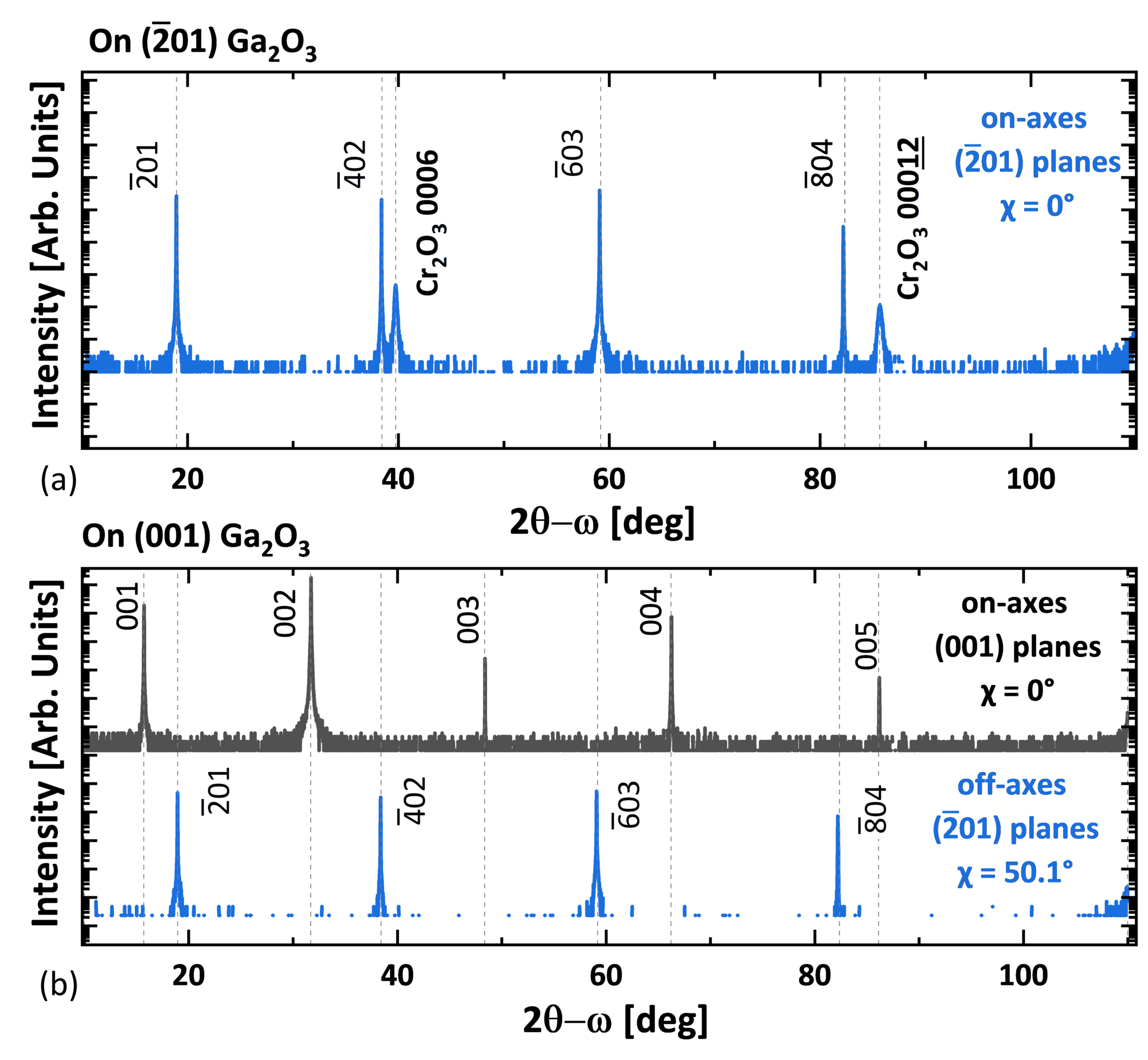}
    \caption{\label{fig:FigS1} In (a), symmetric $2\theta$--$\omega$ scans of the Cr$_2$O$_3$ epilayer grown on $(\bar{2}01)$ Ga$_2$O$_3$, acquired at $\chi = 0^\circ$ to probe the out-of-plane (out-of-plae) scattering vectors. In (b), symmetric $2\theta$--$\omega$ scans of the Cr$_2$O$_3$ epilayer grown on $(001)$ Ga$_2$O$_3$, acquired at $\chi = 0^\circ$ (top) and $\chi = 50.1^\circ$ (bottom) to investigate the on-axis $(001)$ and off-axis $(\bar{2}01)$ Ga$_2$O$_3$ planes, respectively. Each peak is labeled, and vertical dashed lines indicate the nominal $2\theta$ positions of the substrate and epilayer reflections.}
\end{figure*}

The epitaxial relationship study conducted throughout this work highlights the complexity of the Ga$_2$O$_3$ substrate and the influence of its low-symmetry monoclinic unit cell on epilayer growth. In particular, the (001) oriented Ga$_2$O$_3$, despite being the most applied for devices fabrication, poses serious challenges when trying to characterize epilayers grown on its surface, \textit{i.e.}, no epilayer planes are found to be parallel to the substrate $(001)$ planes, making the identification of the epitaxial relationship less straightforward with respect to epilayers grown on different orientations of Ga$_2$O$_3$, \textit{e.g.}, ($\bar{2}$01) (see Figure \ref{fig:FigS1}(a,b)). Only by combining two different X-ray diffractometer, the epitaxial orientation of Cr$_2$O$_3$/(001) Ga$_2$O$_3$ is detected (as reported in the Main Manuscript - Figure 1). The experimental finding is Cr$_2$O$_3$ (0001) || Ga$_2$O$_3$ (101). The available data so far do not allow us to unambiguously assign the in-plane orientation.\\

\begin{figure*}
    \centering
    \includegraphics[width=\textwidth]{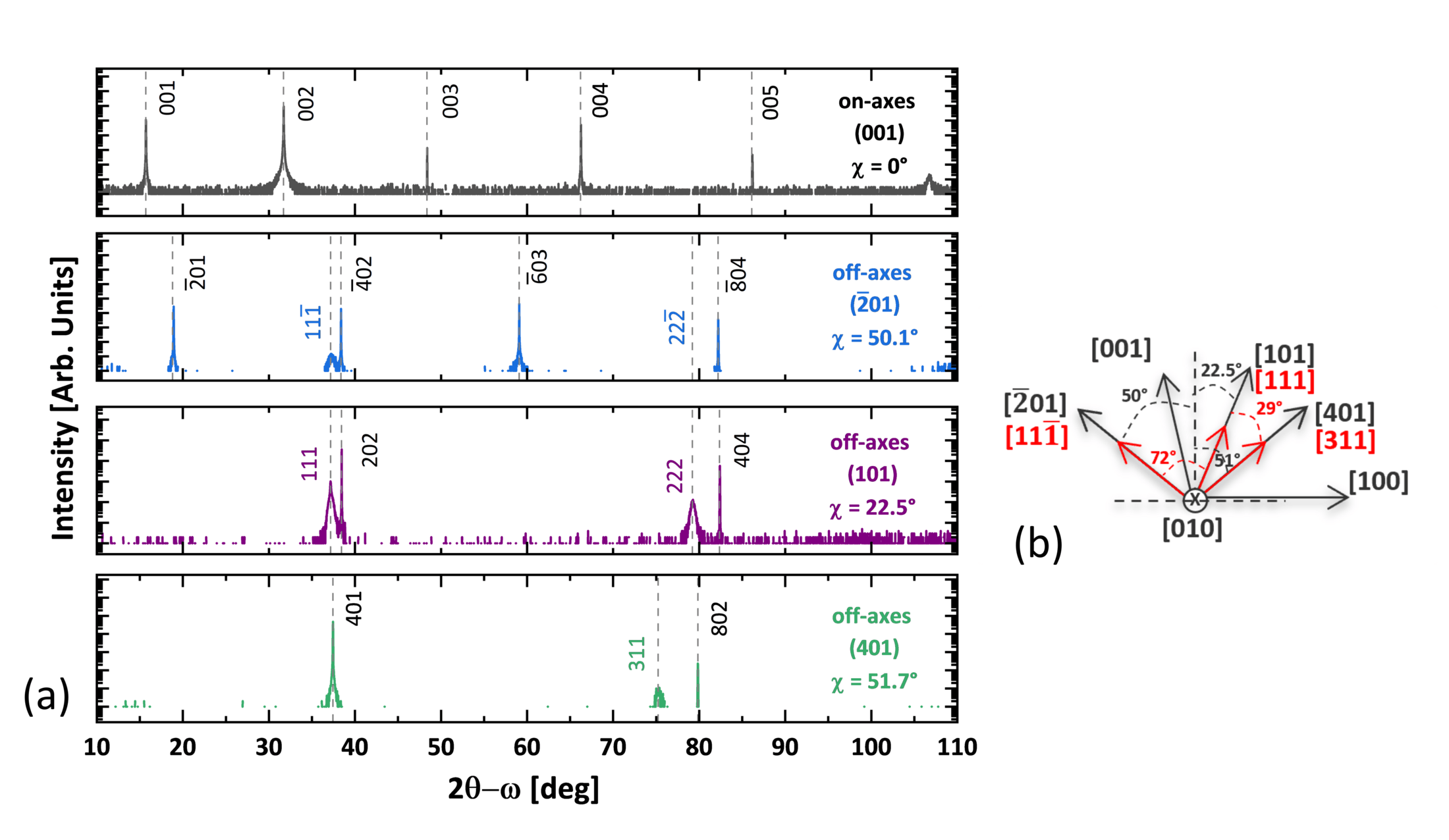}
    \caption{\label{fig:FigS2} In (a) XRD measurements for a NiO/(001) Ga$_2$O$_3$ sample. From top to bottom, 2$\theta$-$\omega$ scans for (001), ($\bar{2}$01), (101) and (401) Ga$_2$O$_3$ reflections. This allowed proper characterization of NiO/(001) Ga$_2$O$_3$ epitaxial relationship and allowed to build the schematic vector alignment reported (b)}
\end{figure*}

The investigation approach adopted to characterize Cr$_2$O$_3$/(001) Ga$_2$O$_3$ was subsequently extended to the NiO/(001) Ga$_2$O$_3$ system, for which the epitaxial relationship is not straightforward, as reported in the literature.\cite{nakagomi_crystal_2020} Also in this system, we successfully identified the NiO (111) planes as parallel to the $(101)$ planes of Ga$_2$O$_3$. Furthermore, additional relevant Ga$_2$O$_3$ planes, namely $(\bar{2}01)$ and (401), were probed, providing further insight into the relative alignment of the NiO epilayer. The results are summarized in Figure \ref{fig:FigS2}, where the individual symmetric 2$\theta$-$\omega$ scans are shown with the substrate and epilayer reflections labeled. The corresponding vector-alignment construction is also included to provide a direct visualization of the relative orientation between the substrate and epilayer, with substrate and epilayer vectors represented in black and red, respectively.\\

\begin{figure*}    
   \centering
    \includegraphics[width=\textwidth]{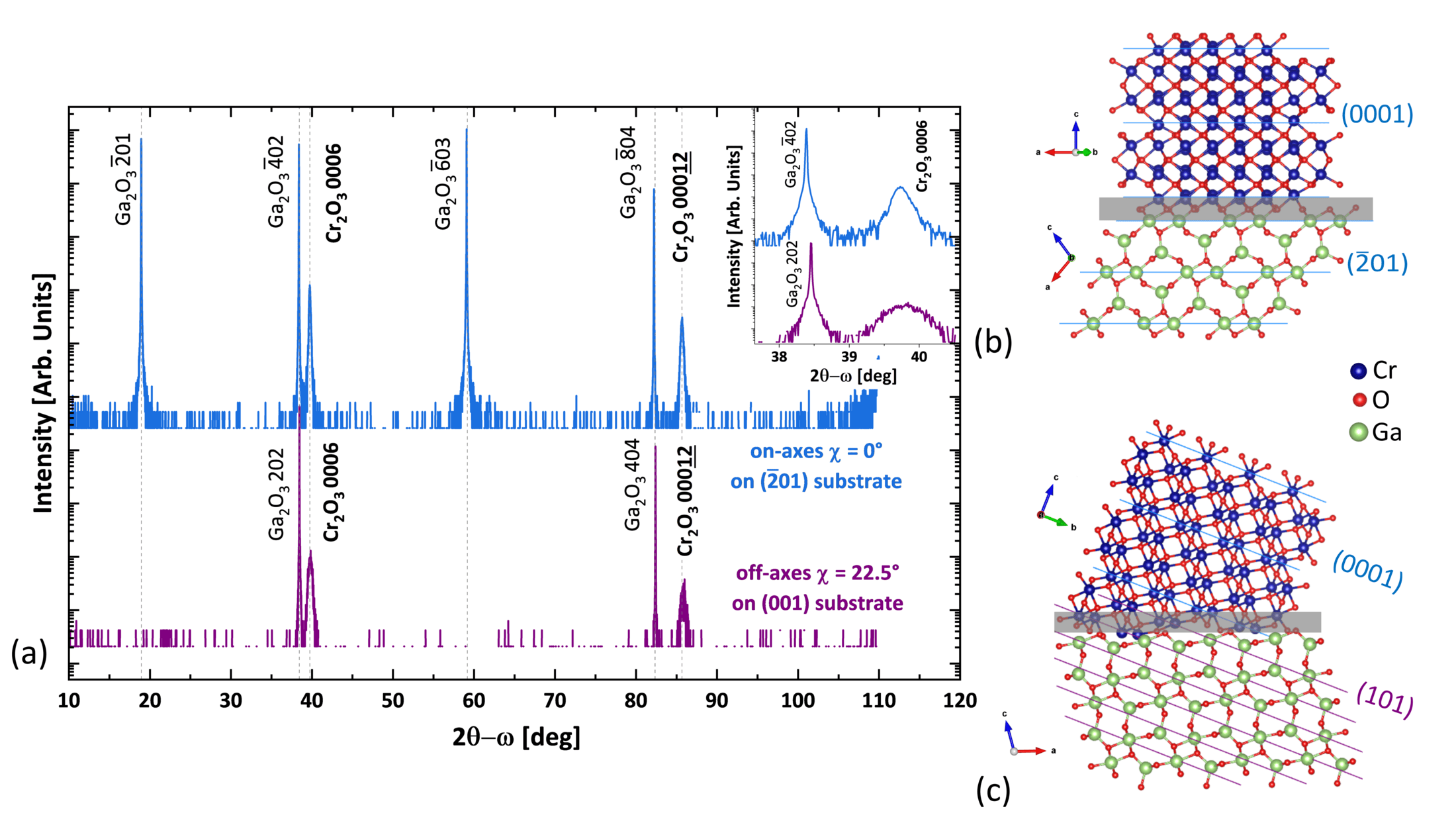}
    \caption{\label{fig:FigS3} In (a) symmetric $2\theta$-$\omega$ scans of Cr$_2$O$_3$ grown on $(\bar{2}01)$- and (001)-oriented Ga$_2$O$_3$, shown in blue and purple, respectively. Measurements were acquired using a Rigaku SmartLab X-ray diffractometer at $\chi = 0^\circ$ and $\chi = 22.5^\circ$, to probe the on-axis $(\bar{2}01)$ and off-axis (101) Ga$_2$O$_3$ planes. Each peak of both substrate and epilayer are labeled and the vertical grey dashed lines represent the nominal position expected for each peak. The inset shows a magnified view of the Cr$_2$O$_3$ 0006 and the corresponding substrate reflection, highlighting the small angular separation between the two substrate peaks. In (b) and (c) schematic representation, generated using VESTA, of the epitaxial relationships of Cr$_2$O$_3$ grown on ($\bar{2}$01) and (001) oriented Ga$_2$O$_3$ substrates. The dual alignment of the Cr$_2$O$_3$ (0001) planes, depending on the substrate orientation, is highlighted. The grey boxes are superimposed at the interface where the exact atomic arrangement is not defined. The representation is in cross-view, \textit{i.e.}, out-of-plane substrate vector aligned vertically.}
\end{figure*}

\begin{figure*}
    \centering
    \includegraphics[width=\textwidth]{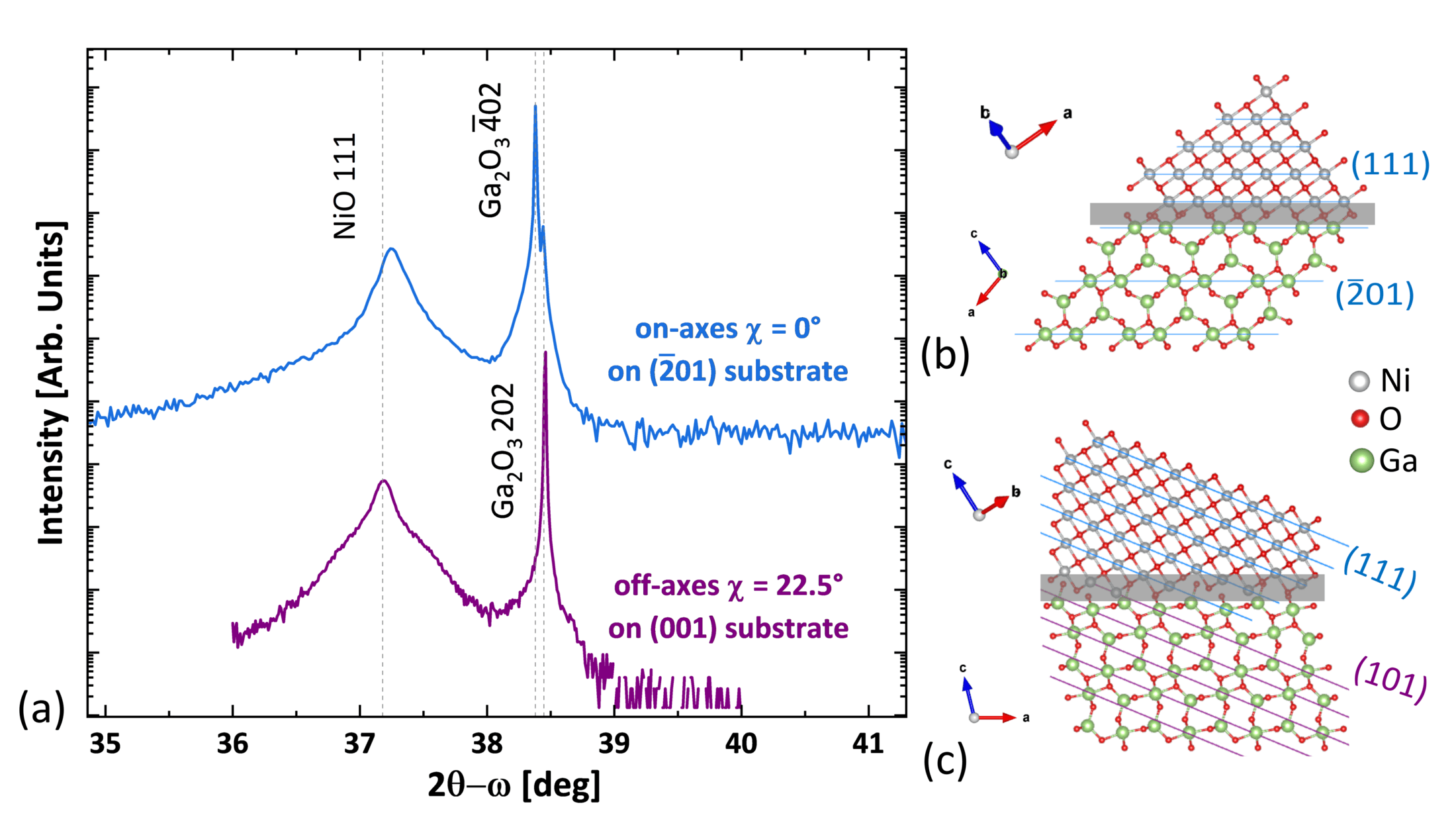}
    \caption{\label{fig:FigS4} In (a) symmetric 2$\theta$-$\omega$ scans for NiO on ($\bar{2}$01) and (001) out-of-plane oriented Ga$_2$O$_3$ substrate, in blue and purple respectively. Each peak of both substrate and epilayer are labeled and the vertical grey dashed lines represent the nominal position expected for each peak. Measurements were acquired using a Rigaku SmartLab X-ray diffractometer at $\chi = 0^\circ$ and $\chi = 22.5^\circ$, to probe the on-axis $(\bar{2}01)$ and off-axis (101) Ga$_2$O$_3$ planes. The two scans reveal the two different epitaxial relationships of NiO (111) on Ga$_2$O$_3$ according to substrate nominal out-of-plane orientation. In (b) and (c) schematic representation, generated using VESTA, of the epitaxial relationships of NiO grown on ($\bar{2}$01) and (001) oriented Ga$_2$O$_3$ substrates. The dual alignment of the NiO (111) planes, depending on the substrate orientation, is highlighted. The grey boxes are superimposed at the interface where the exact atomic arrangement is not defined. The representation is in cross-view, \textit{i.e.}, out-of-plane substrate vector aligned vertically.} 
\end{figure*}

The experimental results obtained for the two heterostructures highlight that the two substrate orientations investigated, \textit{i.e.}, $(001)$ and $(\bar{2}01)$, lead to distinct epitaxial alignments for the same crystallographic plane families of Cr$_2$O$_3$ and NiO, namely, $(0001)$ and $(111)$, respectively. These results are summarized in Figures~\ref{fig:FigS3} and~\ref{fig:FigS4}, which report the corresponding 2$\theta$-$\omega$ scans together with VESTA schematics of the epilayer/substrate stacks.\\

\begin{figure*}
    \centering
   \includegraphics[width=\textwidth]{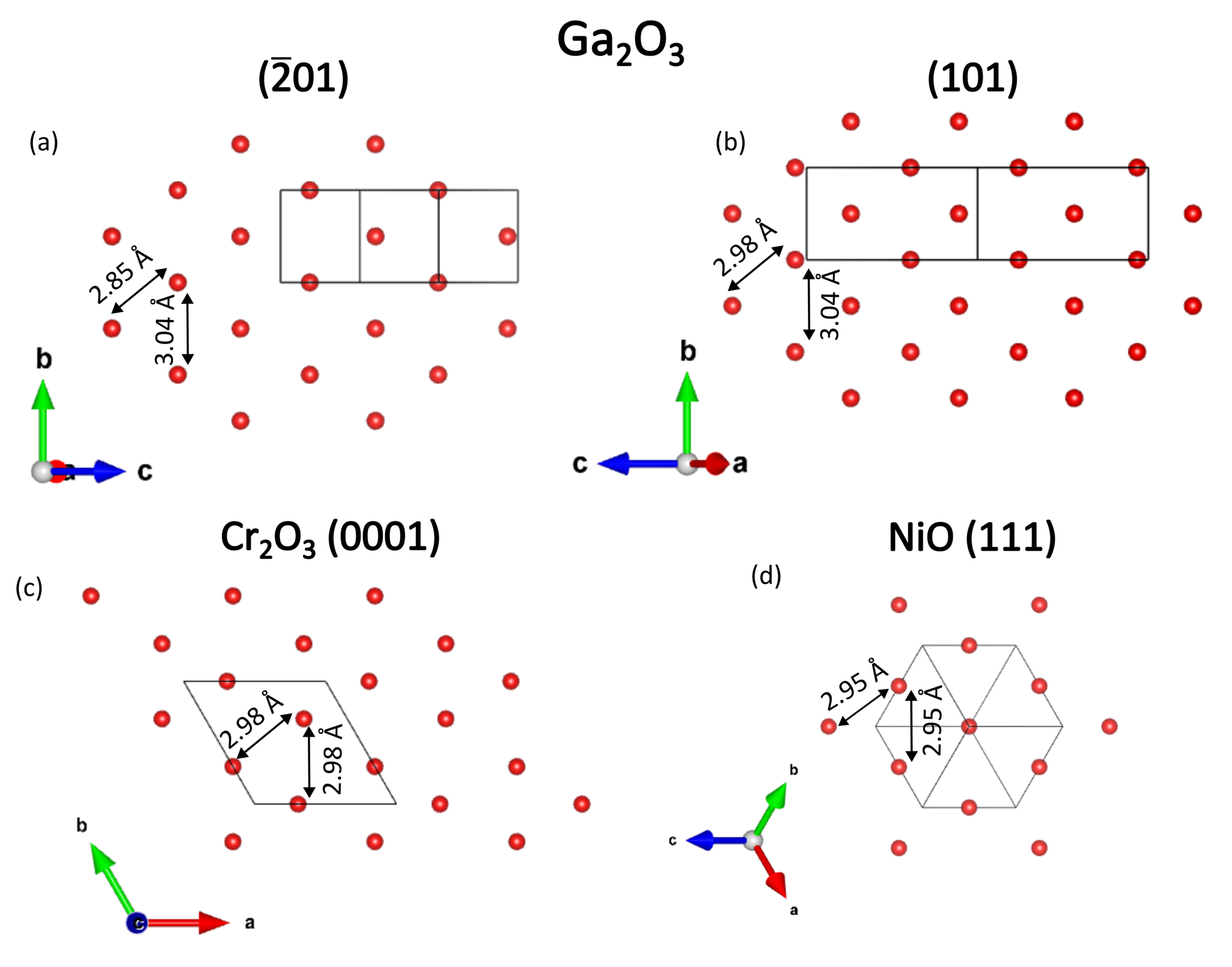}
    \caption{\label{fig:FigS5} Oxygen sublattices of the Ga$_2$O$_3$ $(\bar{2}01)$ and $(101)$ plane are shown in (a) and (b), respectively, from top-view, with selected interatomic O-O distances reported. In (c,d), the oxygen sublattice of the Cr$_2$O$_3$ $(0001)$ and NiO (111) planes is shown to highlight the correspondence between the oxygen sublattices of the substrate and epilayers.} 
\end{figure*}

To understand why the $(\bar{2}01)$ and $(101)$ planes are involved in the epitaxial relationship, we investigated the corresponding oxygen sublattices and found them to be crystallographically equivalent, as reported in Figure \ref{fig:FigS5}(a,b). Furthermore,  the O-O interatomic distances of these two planes are comparable with the one of (0001) Cr$_2$O$_3$  and (111) NiO planes (see Figure \ref{fig:FigS5}(c,d)). The oxygen sublattice is well known to play a key role in determining epitaxial compatibility between different crystal structures. Furthermore, a recent study by Oshima et al.~\cite{oshima_mapping_2026} demonstrated that these two planes are equivalent when the monoclinic unit cell of $\mathrm{Ga_2O_3}$ is represented in terms of its higher-symmetry cubic oxygen sublattice. This observation provides further support for our findings and offers a structural explanation for the two possible epitaxial alignments of the investigated \textit{p}-type oxides.\\

\section{Calculation method}
First-principles density functional theory (DFT) calculations were performed using the Vienna Ab initio Simulation Package (VASP)\cite{kresse_PRB_1999} to determine the lattice parameters of bulk Ga$_2$O$_3$ and Cr$_2$O$_3$. To ensure high accuracy, the plane-wave energy cutoff was set to 340 eV for Ga$_2$O$_3$ and NiO, and Cr$_2$O$_3$. The Brillouin zone was sampled using a $\Gamma$-centered $k$-point grid with an $R_k$ value of 20 for all bulk calculations, which was rigorously tested for energy convergence. The converged structures were then used to construct the surfaces, which were subsequently mapped as inputs for our interface modeling.\\
\begin{table}[h!]
\begin{center}
\caption{Chemical potential ($\mu$) values used for surface construction and interface structure modeling.}
\begin{tabular}{ c c c c c } 
\hline
 & \makecell{O-rich\\$\beta$-Ga$_{2}$O$_{3}$} & \makecell{O-poor\\$\beta$-Ga$_{2}$O$_{3}$ } & \makecell{O-rich\\Cr$_{2}$O$_{3}$ } & \makecell{O-poor\\Cr$_{2}$O$_{3}$ }\\
\hline
$\mu_{\text{M}}$ (eV) & -7.995 & -2.37 & -12.395 & -5.315 \\ 
$\mu_{\text{O}}$ (eV) & -4.76 & -8.51 & -5.191 & -8.734 \\ 
\hline
\end{tabular}
\label{tab:chem_potential}
\end{center}
\end{table}\\

\begin{figure}
    \centering
     \includegraphics[width=\columnwidth]{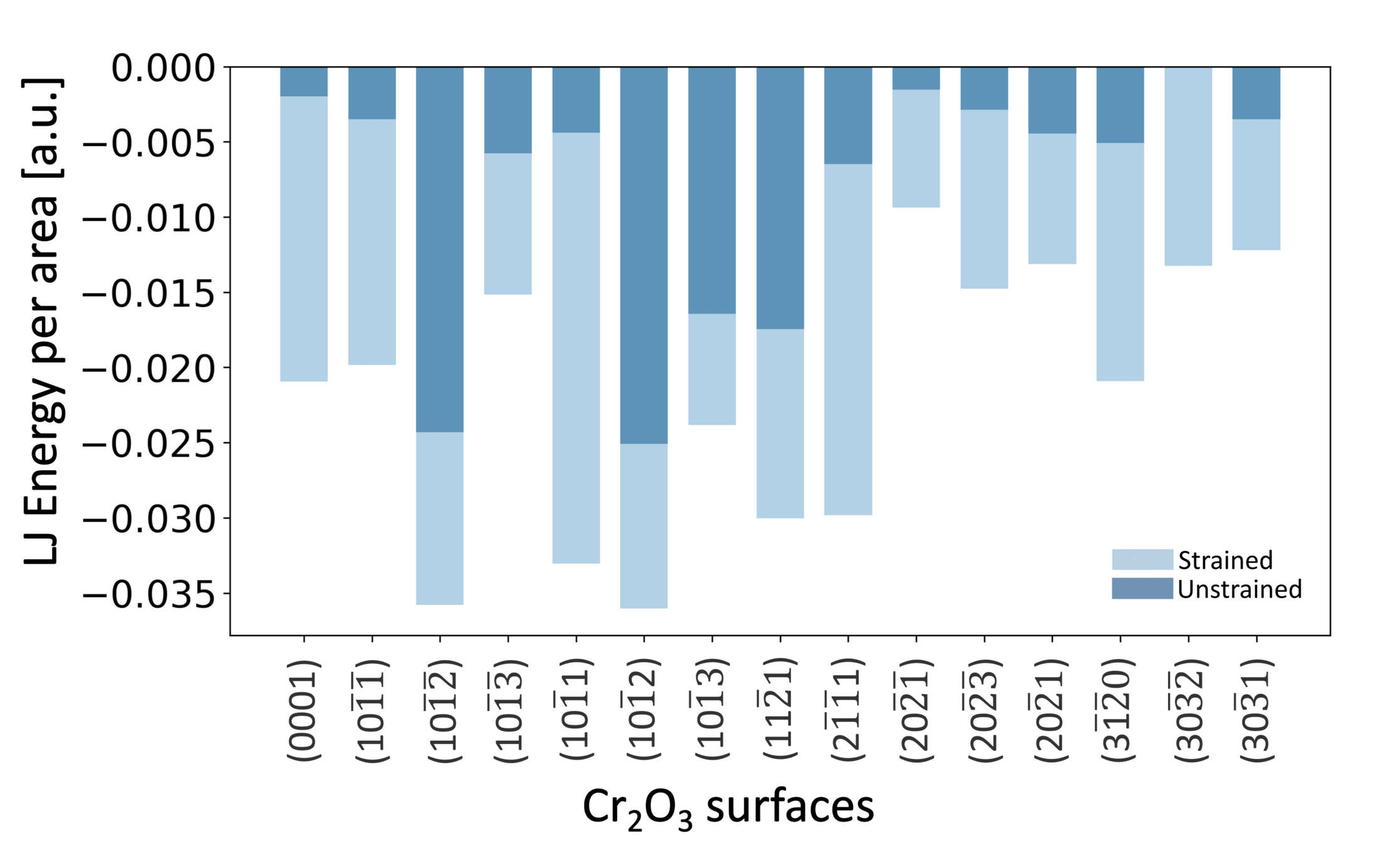}
    \caption{\label{fig:FigS6} Evaluated Lennard-Jones (LJ) energy per unit area for O-poor terminated Ga$_{2}$O$_{3}$($\bar{2}$01)/Cr$_{2}$O$_{3}$ ($hkl$) interfaces with $-3 \leq h,k,l \leq 3$ and a maximum allowable area strain of 8\%. A lower LJ energy per unit area indicates better interface matching. Light and dark blue bars represent the goodness of match with and without strain, respectively.} 
\end{figure}

\begin{figure}
    \centering
     \includegraphics[width=\columnwidth]{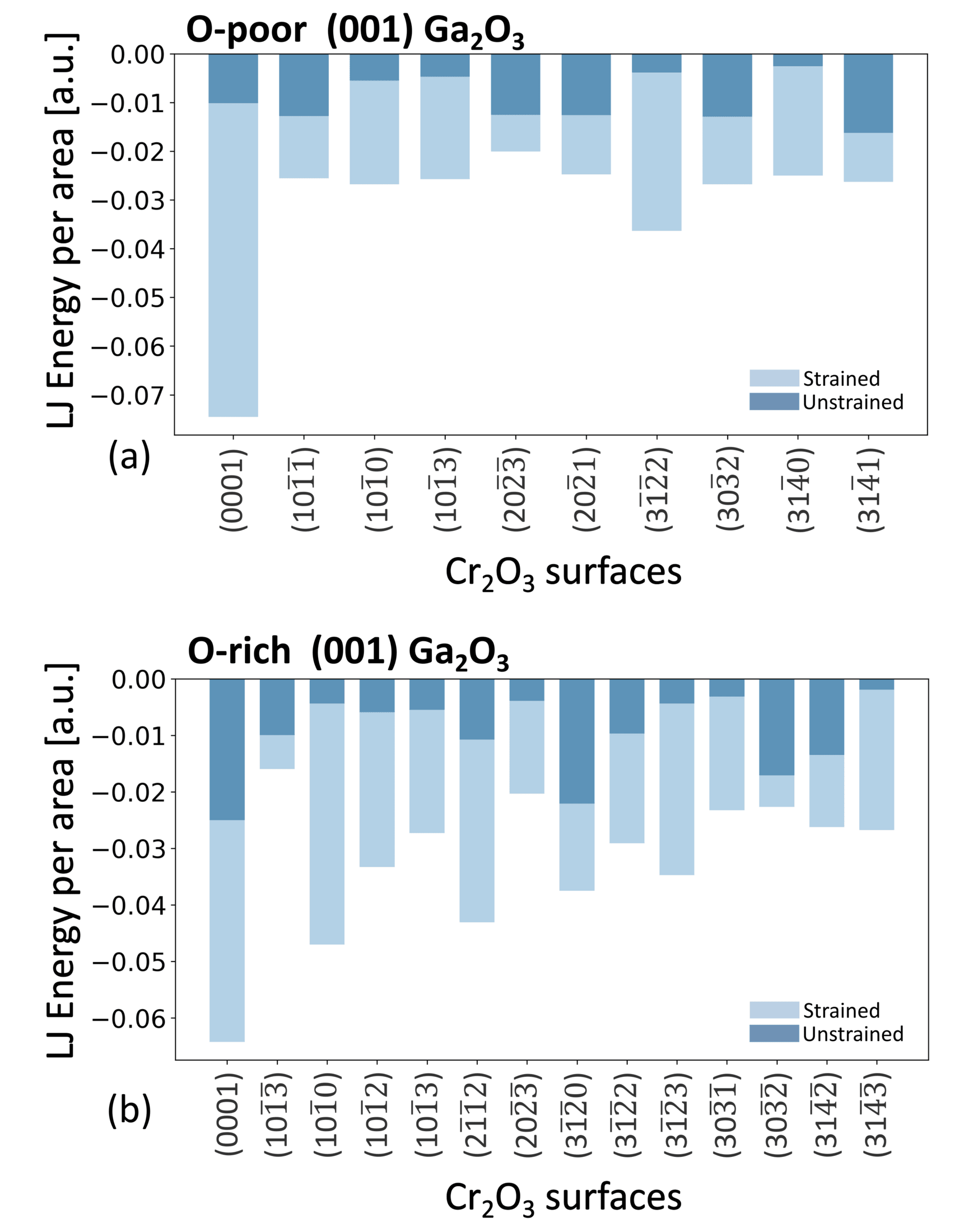}
    \caption{\label{fig:FigS7}In (a) and (b) evaluated Lennard-Jones (LJ) energy per unit area for O-poor and O-rich terminated Ga$_{2}$O$_{3}$(001)/Cr$_{2}$O$_{3}$ ($hkl$) interfaces with $-3 \leq h,k,l \leq 3$ and a maximum allowable area strain of 8\%. A lower LJ energy per unit area indicates better interface matching. Light and dark blue bars represent the goodness of match with and without strain, respectively.} 
\end{figure}

In Figure \ref{fig:FigS6}, we report the evaluated LJ energy per unit area for the O-poor terminated Ga$_2$O$_3$($\bar{2}$01)/Cr$_2$O$_3$ interfaces. The LJ energies are overall higher with respect to the O-rich terminated Ga$_2$O$_3$($\bar{2}$01) surface (see Main Manuscript - Figure 4) and for this reason not representing the most favorable interfacial match.
In Figure \ref{fig:FigS7}(a,b), we report the LJ energy per unit area calculated for O-poor and O-rich terminated Ga$_2$O$_3$(001)/Cr$_2$O$_3$ interfaces. For both substrate surface terminations, the match with Cr$_2$O$_3$ (0001) interface is displaying the absolute lowest LJ energy. However, this match is not correctly representing the experimental findings due to the challenges in detecting low-indexes Cr$_2$O$_3$ planes aligned parallel to the (001) Ga$_2$O$_3$. Considering higher-index Cr$_2$O$_3$ planes could provide additional candidate configurations; however, this would
substantially increase the computational cost and complexity
of the interface modeling. \\


\section{References}

\bibliography{Reference}